\documentclass[%
paper=letter,%
pagesize,%
numbers=noendperiod,%
captions=nooneline,%
abstracton,%
DIV=14
]{scrartcl}

\usepackage[T1]{fontenc}%
\usepackage{lmodern}%
\usepackage[american]{babel}%
\usepackage{microtype}%

\usepackage[
hyperref,%
table%
]{xcolor}%

\usepackage{scrlayer-scrpage}%

\usepackage{eso-pic}%
\usepackage{rotating}%
\usepackage{amsmath}%
\usepackage{mathtools}%
\usepackage{amssymb}%
\usepackage{amsthm}%
\usepackage{etoolbox}%
\usepackage{bm}%
\usepackage{bbm}%
\usepackage{enumitem}%
\usepackage{graphicx}%
\usepackage{siunitx} %
\usepackage{grffile}%
\usepackage{tikz}%
\usetikzlibrary{shapes, decorations.pathreplacing}
\usepackage{wrapfig}%
\usepackage{tabularx}%
\usepackage{bbm}
\usepackage{siunitx}%
\usepackage{booktabs}%
\usepackage{multirow}%
\usepackage{vruler}%
\usepackage{fancyvrb}%
\usepackage{listings}%
\usepackage{csquotes}%
\usepackage[
style=authoryear,%
dashed=false,%
hyperref=true,%
useprefix=true,%
maxnames=2,%
maxbibnames=6,%
uniquename=false,%
seconds=true,%
date=iso,%
urldate=iso%
]{biblatex}%
\usepackage[
hypertexnames=false,%
setpagesize=false,%
pdfborder={0 0 0},%
pdfstartview=Fit,%
bookmarksopen=true,%
bookmarksnumbered=true%
]{hyperref}%
\xdefinecolor{lightgray}{RGB}{247, 247, 247}%
\xdefinecolor{semilightgray}{RGB}{240, 240, 240}%
\xdefinecolor{middlegray}{RGB}{127, 127, 127}%
\xdefinecolor{blue}{RGB}{58, 95, 205}%
\xdefinecolor{deepskyblue}{RGB}{0, 154, 205}%
\xdefinecolor{chocolate}{RGB}{205, 102, 29}%

\setkomafont{pageheadfoot}{\normalfont\normalcolor\sffamily}%
\setkomafont{pagenumber}{\normalfont\normalcolor\sffamily}%
\automark{section}%
\setkomafont{captionlabel}{\normalfont\normalcolor\sffamily\bfseries}%

\setlist{%
  align=left,%
  labelsep=*,%
  leftmargin=*,%
  topsep=1mm,%
  itemsep=0mm%
}
\newcommand*{\mysquare}{\rule[0.18em]{0.36em}{0.36em}}
\newcommand*{\mytriangle}{\raisebox{0.12em}{\resizebox{0.48em}{0.48em}{$\blacktriangleright$}}}
\newcommand*{\mybar}{\rule[0.32em]{0.62em}{0.08em}}
\newcommand*{\mydot}{\raisebox{0.14em}{\resizebox{0.44em}{!}{$\bullet$}}}
\setlist[itemize,1]{label={\mysquare\ }}%
\setlist[itemize,2]{label={\mytriangle\ }}%
\setlist[itemize,3]{label={\mybar\ }}%
\setlist[itemize,4]{label={\mydot\ }}%
\setlist[enumerate,1]{label=\arabic*)}%
\setlist[enumerate,2]{label=\arabic{enumi}.\arabic*)}%
\setlist[enumerate,3]{label=\arabic{enumi}.\arabic{enumii}.\arabic*)}%

\makeatletter
\newcommand\myisodate{\number\year-\ifcase\month\or 01\or 02\or 03\or 04\or 05\or 06\or 07\or 08\or 09\or 10\or 11\or 12\fi-\ifcase\day\or 01\or 02\or 03\or 04\or 05\or 06\or 07\or 08\or 09\or 10\or 11\or 12\or 13\or 14\or 15\or 16\or 17\or 18\or 19\or 20\or 21\or 22\or 23\or 24\or 25\or 26\or 27\or 28\or 29\or 30\or 31\fi}%
\makeatother
\newcommand*{\abstractnoindent}{}%
\let\abstractnoindent\abstract
\renewcommand*{\abstract}{\let\quotation\quote\let\endquotation\endquote
  \abstractnoindent}
\deffootnote[1em]{1em}{1em}{\textsuperscript{\thefootnotemark}}%
\lstdefinestyle{input}{
  backgroundcolor=\color{semilightgray},%
  commentstyle=\itshape\color{chocolate},%
  keywordstyle=\color{blue},%
  stringstyle=\color{blue},%
  numbers=left,%
  numbersep=4.8pt,%
  numberstyle=\color{darkgray!80}\tiny%
}
\lstdefinestyle{output}{
  backgroundcolor=\color{lightgray}%
}

\lstdefinestyle{Lstyle}{
  language=[LaTeX]TeX,%
  texcs={},%
  otherkeywords={}%
}

\lstnewenvironment{Linput}[1][]{%
  \lstset{style=input, style=Lstyle}
  #1%
}{\vspace{-0.25\baselineskip}}%

\lstnewenvironment{Loutput}[1][]{%
  \lstset{style=output, style=Lstyle}
  #1%
}{\vspace{-0.25\baselineskip}}%

\lstdefinestyle{Rstyle}{
  language=R,%
  keywords={if, else, repeat, while, function, for, in, next, break},%
  otherkeywords={}%
}

\lstnewenvironment{Rinput}[1][]{%
  \lstset{style=input, style=Rstyle}
  #1%
}{\vspace{-0.25\baselineskip}}%

\lstnewenvironment{Routput}[1][]{%
  \lstset{style=output, style=Rstyle}
  #1%
}{\vspace{-0.25\baselineskip}}%

\DeclareNameAlias{sortname}{last-first}%
\DefineBibliographyExtras{american}{\DeclareQuotePunctuation{}}%
\renewbibmacro*{volume+number+eid}{%
  \setunit*{\addcomma\space}%
  \printfield{volume}%
  \printfield{number}}
\DeclareFieldFormat*{number}{(#1)}
\DeclareFieldFormat*{title}{#1}%
\DeclareFieldFormat{doi}{%
  \ifhyperref
  {\href{http://dx.doi.org/#1}{\nolinkurl{doi:#1}}}%
  {\nolinkurl{doi:#1}}}%
\renewbibmacro*{in:}{}%
\DeclareFieldFormat{isbn}{ISBN #1}%
\DeclareFieldFormat{pages}{#1}%
\DeclareFieldFormat{url}{\url{#1}}%
\DeclareFieldFormat{urldate}{\mkbibparens{#1}}%
\renewcommand*{\cite}[2][]{\textcite[#1]{#2}}%

\newif\ifstarttheorem
\newtheoremstyle{mythmstyle}%
{0.5em}%
{0.5em}%
{}%
{}%
{\sffamily\bfseries\global\starttheoremtrue}%
{}%
{\newline}%
{\thmname{#1}\ \thmnumber{#2}\ \thmnote{(#3)}}%
\theoremstyle{mythmstyle}%
\newtheorem{definition}{Definition}[section]%

\newtheorem{remark}[definition]{Remark}

\newtheorem{algorithm}[definition]{Algorithm}

\makeatletter
\preto\itemize{%
  \if@inlabel
  \ifstarttheorem
  \mbox{}\par\nobreak\vskip\glueexpr-\parskip-\baselineskip+0.25em\relax\hrule\@height\z@
  \fi%
  \fi%
  \global\starttheoremfalse%
  \def\tempa{proof}%
  \ifx\tempa\mycurrenvir
  \ifstarttheorem
  \mbox{}\par\nobreak\vskip\glueexpr-\parskip-\baselineskip+0.25em\relax\hrule\@height\z@
  \fi%
  \fi%
  \global\starttheoremfalse%
}
\preto\enditemize{\global\starttheoremfalse}
\makeatother

\makeatletter
\preto\enumerate{%
  \if@inlabel
  \ifstarttheorem
  \mbox{}\par\nobreak\vskip\glueexpr-\parskip-\baselineskip+0.25em\relax\hrule\@height\z@
  \fi%
  \fi%
  \global\starttheoremfalse%
  \def\tempa{proof}%
  \ifx\tempa\mycurrenvir
  \ifstarttheorem
  \mbox{}\par\nobreak\vskip\glueexpr-\parskip-\baselineskip+0.25em\relax\hrule\@height\z@
  \fi%
  \fi%
  \global\starttheoremfalse%
}
\preto\endenumerate{\global\starttheoremfalse}
\makeatother

\newcommand{\ou}[3]{%
  \mathrel{%
    \vcenter{\offinterlineskip
      \ialign{##\cr$#1$\cr\noalign{\kern-#3}$#2$\cr}%
    }%
  }%
}
\newcommand*{\omu}[3]{\underset{#3}{\overset{#1}{#2}}}
\newcommand*{\T}{^{\top}}
\renewcommand*{\i}{\leftarrow}
\newcommand*{\isim}{\omu{\text{\tiny{ind.}}}{\sim}{}}

\newcommand*{\IZ}{\mathbbm{Z}}

\newcommand*{\IR}{\mathbbm{R}}

\newcommand*{\U}{\operatorname{U}}

\newcommand*{\N}{\operatorname{N}}

\newcommand*{\ARMA}{\operatorname{ARMA}}
\newcommand*{\GARCH}{\operatorname{GARCH}}

\newcommand*{\I}{\mathbbm{1}}

\newcommand*{\E}{\mathbbm{E}}

\newcommand*{\Var}{\operatorname{Var}}

\newcommand*{\MMD}{\operatorname{MMD}}
\newcommand*{\AVS}{\operatorname{AVS}}
\newcommand*{\AMSE}{\operatorname{AMSE}}
\newcommand*{\AMMD}{\operatorname{AMMD}}

\newcommand*{\R}{\textsf{R}}

\newcommand*{\nbat}{n_{\text{bat}}}
\newcommand*{\nepo}{n_{\text{epo}}}
\newcommand*{\ngen}{n_{\text{gen}}}
\newcommand*{\nkrn}{n_{\text{krn}}}
\newcommand*{\npth}{n_{\text{pth}}}
\newcommand*{\nrep}{n_{\text{rep}}}

\begin{document}
\thispagestyle{plain}
\begin{center}
  \sffamily
  {\bfseries\LARGE Multivariate time-series modeling with generative neural networks\par}
  \bigskip\smallskip
  {\Large Marius Hofert\footnote{Department of Statistics and Actuarial Science, University of
      Waterloo, 200 University Avenue West, Waterloo, ON, N2L
      3G1,
      \href{mailto:marius.hofert@uwaterloo.ca}{\nolinkurl{marius.hofert@uwaterloo.ca}}. The
      author acknowledges support from NSERC (Grant RGPIN-2020-04897).},
    Avinash Prasad\footnote{Department of Statistics and Actuarial Science, University of
      Waterloo, 200 University Avenue West, Waterloo, ON, N2L
      3G1,
      \href{mailto:a2prasad@uwaterloo.ca}{\nolinkurl{a2prasad@uwaterloo.ca}}. The author acknowledges support from NSERC (PGS D Scholarship).},
    Mu Zhu\footnote{Department of Statistics and Actuarial Science, University of
      Waterloo, 200 University Avenue West, Waterloo, ON, N2L
      3G1,
      \href{mailto:mu.zhu@uwaterloo.ca}{\nolinkurl{mu.zhu@uwaterloo.ca}}. The
      author acknowledges support from NSERC (RGPIN-2016-03876).}
    \par\bigskip
    \myisodate\par}
\end{center}
\par\smallskip
\begin{abstract}
  Generative moment matching networks (GMMNs) are introduced as dependence
  models for the joint innovation distribution of multivariate time series
  (MTS).  Following the popular copula--GARCH approach for modeling dependent
  MTS data, a framework based on a GMMN--GARCH approach
  is presented. First, ARMA--GARCH models are utilized to capture the serial
  dependence within each univariate marginal time series. Second, if the number
  of marginal time series is large, principal component analysis (PCA) is used
  as a dimension-reduction step. Last, the remaining cross-sectional dependence
  is modeled via a GMMN, the main contribution of this work. GMMNs are highly flexible and
  easy to simulate from, which is a major advantage over the copula--GARCH
  approach. Applications involving yield curve modeling and the analysis of
  foreign exchange-rate returns demonstrate the utility of the GMMN--GARCH
  approach, especially in terms of producing better empirical predictive
  distributions and making better probabilistic forecasts.
\end{abstract}
\minisec{Keywords}
Generative moment matching networks, learning distributions, copulas, probabilistic forecasts, ARMA--GARCH model, yield curves, exchange-rate dependence.
\minisec{MSC2010}
62H99, 65C60, 60E05, 00A72, 65C10, 62M10. %

\section{Introduction}
\label{sec:intro}
We extend the popular \emph{copula--GARCH approach}
\parencite{jondeaurockinger2006,patton2006} for modeling multivariate time
series (MTS) data by replacing parametric copulas with generative neural
networks. Our primary goal is to make probabilistic forecasts by constructing
empirical predictive distributions. Quantities we are interested in forecasting
include not just the average value but also quantiles, tail probabilities and
others at any given time point in the future.

\subsubsection*{Copula--GARCH approach}
Typically, there are two types of dependence in MTS data: serial dependence
within each univariate time series, and cross-sectional dependence between the
individual time series. A key feature of the copula--GARCH approach is that it
models these two types of dependence separately.

First, for serial dependence within each univariate time series, the
copula--GARCH approach uses a class of generalized auto-regressive conditional
heteroscedasticity (GARCH) models \parencite{bollerslev1986}. While there is a
plethora of literature on univariate time series modeling with a wide range of
models that are tailor-made for capturing various types of serial patterns such
as seasonality, volatility clustering or regime switching, GARCH-type models are
popular choices in the realm of financial econometrics because they are designed
to account for various stylized facts (such as volatility clustering) that are
often present in financial return series data; see
\cite[Chapter~3]{mcneilfreyembrechts2015}.

Second, for cross-sectional dependence between the individual time series, the
copula--GARCH approach leverages Sklar's Theorem \parencite{sklar1959} to allow
the use of any copula to model the joint innovation distribution. This extends
various classic approaches \parencite[e.g.,][]{bollerslev1990, engle2002,
  tse2002} that rely on the multivariate normal distribution to model the
cross-sectional dependence, and is the main attraction that makes the
copula--GARCH approach a highly flexible modeling approach.
For a brief overview of its versatility and popularity in the context of finance and econometrics, see
\cite{patton2012}; for the calibration of these models, see
\cite{aas2016}, \cite{almeida2016}, as well as \cite{oh2017}.

\subsubsection*{Motivation}
While there is a growing collection of copula models to characterize complex
dependence structures, most of them are already limited in moderately large
dimensions and often do not provide an adequate fit to given data
\parencite[e.g.,][]{hofertoldford2018}. Identifying appropriate copula models,
estimating their parameters and, especially, testing their goodness-of-fit and
selecting the final model are time-consuming at best. These modeling steps often
require sophisticated algorithms but still do not lead to an adequate
model among the computationally tractable ones. To address these shortcomings,
we propose to replace parametric copula models in the copula--GARCH approach with a
generative moment matching network (GMMN). In comparison to parametric copulas,
GMMNs can capture a much larger variety of complex dependence structures. We
call this alternative modeling approach the \emph{GMMN--GARCH approach}.

For high-dimensional MTS data, we also incorporate principal component analysis
(PCA) as an intermediate step to reduce the dimensionality; although other
dimension reduction techniques, such as autoencoders, can be used here as well,
we do not focus on this aspect in the current paper. Thus, our proposed
GMMN--GARCH approach consists of the following three components:
\begin{enumerate}
\item\label{framework:timeseries} serial dependence modeling --- while many possibilities can be considered, here we deliberately do not deviate from the copula--GARCH approach and use marginal ARMA--GARCH models;
\item\label{framework:dimred} dimension reduction (optional) --- several choices are available, but when this step is needed we simply apply PCA; and
\item\label{framework:dependence} cross-sectional dependence modeling --- here, the copula--GARCH approach would choose a parametric copula, but we introduce the use of GMMNs, the main contribution of this paper.
\end{enumerate}

\subsubsection*{Related use of neural networks}
Neural networks have appeared in the time series literature before.
For example, quite a few research groups have used neural networks to replace the aforementioned GARCH model for serial dependence marginally within each individual time series \parencite[e.g.,][]{LuaQueCao2016, LiuSo2020}.
Another common use of neural networks for time series data is to model not only the serial dependence marginally within each individual time series, but also their conditional relationships with multiple covariates, which themselves may be time series, too \parencite[e.g.,][]{chaudhuri2016, yu2020}.
By contrast, we use neural networks to replace the copula model for cross-sectional dependence jointly between individual time series,
and make improved probabilistic forecasts as a result.
We are not yet aware of such attempts in the literature.

Neural networks also have been used by, for example, \cite{ling2020} in the context of copula modeling. Our work differs from theirs in two ways. First, they learn their copulas in a static setting, not in the context of MTS data. Second, they restrict themselves to Archimedean copulas (which are copulas representable as $C(u_1,\dots,u_d)=\psi(\psi^{-1}(u_1)+\dots+\psi^{-1}(u_d))$
with certain conditions on $\psi$), and only use neural networks to learn the underlying generator (the function $\psi$). While a nonparametric generator can give considerable flexibility to the corresponding Archimedean copula, it is not fully flexible as clearly not all dependence structures are Archimedean.

\subsubsection*{Organization}
In Section~\ref{sec:framework}, we outline our framework for modeling MTS
data. In particular, we focus on the novel integration of GMMNs within this
framework. In Section~\ref{sec:probforecast}, we discuss how to make probabilistic forecasts and evaluate the performance of our MTS model. Then in Section~\ref{sec:applications}, we showcase our GMMN-based
multivariate time series model in applications to yield curve and exchange-rate
data.  Section~\ref{sec:concl} provides concluding remarks.

\section{Framework for multivariate time series modeling}\label{sec:framework}
Let $(\bm{X}_t)_{t \in \IZ}$ denote a $d$-dimensional time series of interest,
where $\bm{X}_t=(X_{t,1},\dots,X_{t,d})$. Furthermore, consider a stretch of
$\tau$ realizations from $(\bm{X}_t)_{t \in \IZ}$ denoted by
$\bm{X}_1,\dots,\bm{X}_{\tau}$. For applications in finance, these are often
log-returns of $d$ asset prices (and negative log-returns for
applications in risk management); see Section~\ref{sec:applications} for more details.
In this section, we describe in more depth the three modeling components outlined in Section~\ref{sec:intro} above.

\subsection{Modeling serial dependence}\label{sec:serial}

The \emph{ARMA--GARCH models} in Step~\ref{framework:timeseries} are
ARMA models with GARCH errors; see \cite[Section~4.2.3]{mcneilfreyembrechts2015}.
An $\ARMA(p_{1j},q_{1j})$--$\GARCH(p_{2j},q_{2j})$ model has the form
\begin{align*}
  &X_{t,j} =\mu_{t,j} + \sigma_{t,j}Z_{t,j},\\
  &\mu_{t,j}=\mu_j+\sum_{k=1}^{p_{1j}}\phi_{jk}(X_{t-k,j}-\mu_{j}) + \sum_{l=1}^{q_{1j}}\gamma_{jl}(X_{t-l,j}-\mu_{t-l,j}),\\
  &\sigma^2_{t,j}=\omega_{j}+\sum_{k=1}^{p_{2j}}\alpha_{jk}(X_{t-k,j}-\mu_{t-k,j})^2+\sum_{l=1}^{q_{2j}}\beta_{jl}\sigma^2_{t-l,j},
\end{align*} %
where, for each component $j=1,\dots,d$, one has $\mu_j\in \IR$, $\omega_{j}> 0$, and
$\alpha_{jk}, \beta_{jl}\geq 0$ for all $k, l$. Additional conditions on the coefficients $\phi_{jk}$, $\gamma_{jl}$, $\alpha_{jk}$ and $\beta_{jl}$ are necessary to ensure that the $\ARMA$--$\GARCH$ processes are causal and covariance stationary; see, e.g.,
\cite[Section~4.1.2--4.2.2]{mcneilfreyembrechts2015} for the details. For each $j=1,\dots,d$, the \emph{innovations} $Z_{t,j}$ in the definition
of the ARMA--GARCH model are independent and identically distributed (iid)
random variables with $\E(Z_{t,j})=0$ and $\Var(Z_{t,j})=1$; their realizations
after fitting marginal $\ARMA(p_{1j},q_{1j})$--$\GARCH(p_{2j},q_{2j})$ models are known as
\emph{standardized residuals} and denoted by $\hat{Z}_{t,j}$,
$t=1,\dots,\tau$ and $j=1,\dots,d$. In financial time series
applications, common choices of innovation distributions include the standard
normal, the scaled $t$ and the skewed $t$ distribution.

Fitting marginal time series models is typically done by fitting low-order
models with likelihood-based methods and selecting the most adequate fit using
the AIC/BIC model selection criterion among the candidate models.  A popular
broad-brush approach is to fit a $\GARCH(1,1)$ model for financial return series --- specifically, an
$\ARMA(0,0)$--$\GARCH(1,1)$ model in our context --- and continue the modeling
based on the standardized residuals $\hat{Z}_{1,j},\dots,\hat{Z}_{\tau,j}$; see
\cite[Chapter~4]{mcneilfreyembrechts2015} or
\cite[Section~6.2.3]{hofertkojadinovicmaechleryan2018}. This procedure is also
referred to as \emph{deGARCHing}. With the help of model diagnostic tools --- for
example, plots of the autocorrelation function (ACF) of
$\hat{Z}_{1,j},\dots,\hat{Z}_{\tau,j}$ and that of
their squared values, Ljung--Box tests or an assessment of the
innovation distribution through Q-Q plots --- one can then assess the
adequacy of each marginal time series model. In what follows we use $\hat{\mu}_{t,j}$ and $\hat{\sigma}_{t,j}^2$ to denote the estimated conditional mean and variance models for the $j$th marginal time series with orders $\hat{p}_{1j},\hat{q}_{1j},\hat{p}_{2j},\hat{q}_{2j}$ and fitted parameters $\hat{\phi}_{jk},\hat{\gamma}_{jl},\hat{\alpha}_{jk},\hat{\beta}_{jl}$.

Having accounted for the marginal serial dependence in this way, the subsequent
analysis in our modeling framework will operate on the standardized residuals
$\hat{\bm{Z}}_t=(\hat{Z}_{t,1},\dots,\hat{Z}_{t,d})$, $t=1,\dots,\tau$,
which are themselves realizations of the innovation random variables, $\bm{Z}_1,\dots,\bm{Z}_{\tau}$, assumed to be iid in the copula--GARCH approach.

Note that any other adequate marginal time series
modeling approach can be
applied in our framework as long as the model's marginal residuals can be considered to be iid realizations
from a continuous distribution. Our choice of ARMA--GARCH models here and in what follows is
motivated only from the fact that these are the most popular marginal time series models used in practice.

\subsection{Dimension reduction}\label{sec:pca}
Two popular dimension-reduction techniques for multivariate financial time
series are factor models and PCA; see
\cite[Chapter~6]{mcneilfreyembrechts2015} and the references therein for a brief
summary.
An approach that is perhaps less discussed in the financial econometrics literature involves using autoencoder neural networks for dimension reduction in which two separate neural network mappings are learned to and from the lower dimensional space;
see \cite{hinton2006}. As dimension reduction is not our main
contribution in this work, we simply utilize PCA in what follows.

Note that PCA is often applied to the MTS data $\bm{X}_1,\dots,\bm{X}_{\tau}$ in the literature; see, e.g.,
\cite{alexander2000}. Apart from reducing the burden of marginal time series modeling, there is
no strong reason why PCA should be applied to potentially non-stationary data.
If dimension reduction is necessary, we find it statistically more sound to apply PCA
to the standardized residuals $\hat{\bm{Z}}_t$ \emph{after} first accounting for any serial
dependence in the marginal time series.

Let $\hat{\Sigma}$ denote the sample covariance matrix of the standardized
residuals $\hat{\bm{Z}}_t$, $t=1,\dots,\tau$. The result from PCA is the
matrix $\hat{\Gamma} \in \IR^{d\times d}$ whose columns consist of the
eigenvectors of $\hat{\Sigma}$, sorted according to decreasing eigenvalues $\hat{\lambda}_1\ge\dots\ge\hat{\lambda}_d\ge 0$. For
the purposes of dimension reduction, $\hat{\bm{Z}}_t$, $t=1,\dots,\tau$,
are transformed to $\hat{\bm{Y}}_t=\hat{\Gamma}_{\cdot,1:k}\T\hat{\bm{Z}}_t$, where
$\hat{\Gamma}_{\cdot,1:k} \in \IR^{d\times k}$ represent the first $k$
columns of $\hat{\Gamma}$ for some $1\le k<d$. As a result, the sample
covariance matrix of $\bm{Y}_t$ is (approximately) diagonal, and the components
of $\bm{Y}_t$ are (approximately) uncorrelated. The $j$th component series $Y_{1,j},\dots,Y_{\tau,j}$, forms realizations of the $j$th principal component,
and the first $k$ principal component series account for
$\sum_{j=1}^k\hat{\lambda}_j/\sum_{j=1}^d\hat{\lambda}_j$ of the total variance.

As dimension reduction is an
optional component in our modeling framework, the next step  involves dependence modeling of either the standardized residuals
$\hat{\bm{Z}}_1,\dots,\hat{\bm{Z}}_{\tau}$ directly or their principal components
$\hat{\bm{Y}}_1,\dots,\hat{\bm{Y}}_{\tau}$. To unify the notation for both cases, we define a $d^{*}$-dimensional time series
$\hat{\bm{Y}}_t=\hat{\Upsilon}\T\hat{\bm{Z}}_t$, where
$\hat{\Upsilon}=\hat{\Gamma}_{\cdot,1:k}$ if dimension reduction is
employed and $\hat{\Upsilon}=I_{d}$ (the
identity matrix in $\IR^{d\times d}$) otherwise; consequently, $d^{*}=k$ in the former case
and $d^{*}=d$ in the latter. Furthermore, we treat $\hat{\bm{Y}}_1,\dots,\hat{\bm{Y}}_{\tau}$ as realizations from $\bm{Y}_t$ and so naturally, $\bm{Y}_t=\Upsilon\T\bm{Z}_t$ with $\Upsilon=\Gamma_{\cdot,1:k}$ if dimension
reduction is used and $\Upsilon=I_d$ otherwise.

\subsection{Modeling cross-sectional dependence}\label{sec:dependence}
The final task in our framework involves the modeling of the iid series
$\bm{Y}_1,\dots,\bm{Y}_{\tau}$. To account for cross-sectional dependence,
we model the joint distribution function $H$ of $\bm{Y}_t$ using Sklar's Theorem as
\begin{align*}
  H(\bm{y})=C(F_1(y_1),\dots,F_{d^{*}}(y_{d^{*}})), \quad \bm{y} \in \IR^{d^{*}},
\end{align*}
where $F_j$, $j=1,\dots,d^{*}$, are the margins of $H$ and
$C:[0,1]^{d^{*}}\rightarrow [0,1]$ is the copula of $(Y_{t,1},\dots,Y_{t,d^{*}})$ for each $t$.

Following a classical copula modeling approach, one first builds the
\emph{pseudo-observations} $\hat{U}_{t,j}=R_{t,j}/(\tau+1)$,
$t=1,\dots,\tau$, $j=1,\dots,d^{*}$, where $R_{t,j}$ denotes the
rank of $\hat{Y}_{t,j}$ among $\hat{Y}_{1,j}\dots,\hat{Y}_{\tau,j}$. The
pseudo-observations are viewed as realizations from $C$ based on which one would
fit candidate copula models; see, e.g.,
\cite[Section~7.5.1]{mcneilfreyembrechts2015} or
\cite[Section~4.1.2]{hofertkojadinovicmaechleryan2018}.
Note that by considering the non-parametric pseudo-observations (even in the case
when we do not apply a dimension reduction technique and thus know the (fitted)
marginal innovation distributions), we reduce the risk of a misspecified margin affecting the estimation of the copula $C$; see
\cite{genest2010} for a theoretical justification of this approach.
Therefore, going forward, we will use the pseudo-observations
$\hat{\bm{U}}_t=(\hat{U}_{t,1},\dots,\hat{U}_{t,d^{*}})$, $t=1,\dots,\tau$,
to model the cross-sectional dependence structure of $\hat{\bm{Y}}_t$.

\subsubsection{Parametric copulas}
A traditional approach for modeling the cross-sectional dependence described by
$\hat{\bm{U}}_1,\dots,\hat{\bm{U}}_{\tau}$ involves the fitting of parametric
copula models, their goodness-of-fit assessment and finally, model selection. There are numerous
families of copula models to consider depending on prominent features of the
dependence structure present in $\hat{\bm{U}}_t$ such as (a)symmetries or a
concentration of points in the lower/upper tail of the joint distribution (or
pairs of such) which hints at an adequate model possessing tail dependence.

A problem with this approach is that it is often hard to find an
adequate copula model for given real-life data, especially in higher dimensions
where typically some pairwise dependencies contradict the corresponding
model-implied marginal copulas; see, e.g.,
\cite{hofertoldford2018}. Another problem is that certain copula models are
computationally expensive to fit and test for goodness-of-fit.  In Section~\ref{sec:applications}, we investigate whether
(the much more flexible) GMMNs can outperform prominent %
elliptical and Archimedean copulas, as well as flexible vine copulas constructed using trees of bivariate copulas \parencite{dissmann2013,aas2009}, in the context of our framework. In what follows we thus shall denote by
$\hat{C}_{\text{PM}}$ a (generic) parametric copula model fitted to the
pseudo-observations $\hat{\bm{U}}_1,\dots,\hat{\bm{U}}_{\tau}$.

\subsubsection{Nonparametric copulas}
Nonparametric copulas, which are more flexible in nature than their parametric counterparts, can also be used to model cross-sectional dependence. A simple and standard nonparametric estimator of $C$ is the empirical copula, which is merely the empirical distribution function of the pseudo-observations $\hat{\bm{U}}_1,\dots,\hat{\bm{U}}_{\tau}$. Since the empirical copula can exhibit large bias when the sample size is small, we also consider a smoothed version of it, known as the empirical beta copula  \parencite{segers2017}, which is a member of the class of empirical Bernstein copulas \parencite{sancetta2004}. The smoothness of the empirical beta copula is a consequence of replacing the indicator functions in the empirical distribution function with a product of various beta distribution functions. Going forward, we use $\hat{C}_{\text{NPM}}$ to denote a (generic) nonparametric estimator of the target copula $C$.

\subsubsection{GMMNs}\label{sec:dependence:GMMN}
We propose to utilize generative neural networks (in particular,
GMMNs) for modeling the cross-sectional dependence structure of the
pseudo-observations $\hat{\bm{U}}_1,\dots,\hat{\bm{U}}_{\tau}$. In our
framework, a generative neural network $f_{\hat{\bm{\theta}}}$ with fitted parameter vector
$\hat{\bm{\theta}}$ is used as an estimator for the distribution of the pseudo-observations. Let
$\hat{C}_{\text{NN}}$ denote the empirical copula based on a sample generated
from a trained GMMN $f_{\hat{\bm{\theta}}}$.

GMMNs, also known as Maximum Mean Discrepancy (MMD) nets, were introduced
simultaneously by \cite{li2015} and \cite{dziugaite2015}. A GMMN
$f_{\bm{\theta}}$ utilizes a kernel maximum mean discrepancy statistic as the
\emph{loss function} to learn the distribution of the
pseudo-observations. Conceptually, $f_{\bm{\theta}}$ can be thought of as a
parametric map from a random vector $\bm{V}_t=(V_{t,1},\dots,V_{t,p})$ with
(known) \emph{prior distribution} $F_{\bm{V}}$ to
$\hat{\bm{U}}_t=(\hat{U}_{t,1},\dots,\hat{U}_{t,d^{*}})$. As is standard in the
literature, we assume that $V_{t,1},\dots,V_{t,p}$ are iid. Typical choices of
$F_{\bm{V}}$ are $\U(0,1)$ or $\N(0,1)$; we utilize the latter. Based on the
fitted GMMN $f_{\hat{\bm{\theta}}}: \IR^p \rightarrow [0,1]^{d^{*}}$ we can then
generate samples with copula $\hat{C}_{\text{NN}}$ as an approximation to
the target copula $C$ of $\hat{\bm{U}}_t$. As demonstrated in \cite{hofertprasadzhu2021}, GMMNs provide a flexible class of models capable of learning a variety of complex dependence structures.

In this paper, we work with a \emph{feedforward neural network} (also known as the multi-layer perceptron), which we simply refer to as \emph{neural network (NN)} in what follows. For details pertaining to the mathematical representation of these NNs, see Appendix~\ref{sec:appendix:GMMN}. Having established the architecture of $f_{\bm{\theta}}$, we will now briefly discuss the loss function and training procedure used for estimating $\bm{\theta}$.

\subsubsection*{Loss function}
To learn $f_{\bm{\theta}}$, we work with $\tau$ training data points consisting
of the pseudo-observations $\hat{\bm{U}}_{1},\dots,\hat{\bm{U}}_{\tau}$. Given
an input sample $\bm{V}_1,\dots,\bm{V}_{\ngen}$ from the prior distribution $F_{\bm{V}}$, the
GMMN generates an output sample $\bm{U}_1,\dots,\bm{U}_{\ngen}$, where
$\bm{U}_t=f_{\bm{\theta}}(\bm{V}_t)$, $t=1,\dots,\ngen$. In selecting an
appropriate loss function, we are naturally interested in measuring whether the two
samples $\hat{U}=(\hat{\bm{U}}_1\T,\dots,\hat{\bm{U}}_{\tau}\T)\T\in [0,1]^{\tau\times
  d^{*}}$ and
$U=(\bm{U}_1\T,\dots,\bm{U}_{\ngen}\T)\T\in[0,1]^{\ngen\times d^{*}}$ can be deemed to come from
the same distribution.

To do so, GMMNs use the \emph{maximum mean discrepancy (MMD)} as loss function, which was introduced as a two-sample test statistic by \cite{gretton2007}. For a given embedding function $\varphi:\mathbb{R}^{d^*} \mapsto \mathbb{R}^{d^\prime}$, the MMD measures the distance between two sample statistics, $(1/\tau)\sum_{t_1=1}^{\tau}\varphi(\hat{\bm{U}}_{t_1})$ and $(1/\ngen)\sum_{t_2=1}^{\ngen}\varphi(\bm{U}_{t_2})$, in the embedded space $\mathbb{R}^{d^\prime}$ via
\begin{align*}
  &\phantom{{}={}}\MMD(\hat{U},U)\\
  &=\Biggl\Vert \frac{1}{\tau}\sum_{t_1=1}^{\tau} \varphi(\hat{\bm{U}}_{t_1})-\frac{1}{\ngen}\sum_{t_2=1}^{\ngen} \varphi(\bm{U}_{t_2}) \Biggr\Vert_2 \\
  &= \Biggl(\frac{1}{\tau^2} \sum_{t_1=1}^{\tau}\sum_{t_2=1}^{\tau}\varphi(\hat{\bm{U}}_{t_1})\T\varphi(\hat{\bm{U}}_{t_2})- \frac{2}{\tau\ngen}\sum_{t_1=1}^{\tau}\sum_{t_2=1}^{\ngen}\varphi(\hat{\bm{U}}_{t_1})\T\varphi(\bm{U}_{t_2}) + \frac{1}{\ngen^2} \sum_{t_1=1}^{\ngen}\sum_{t_2=1}^{\ngen}\varphi(\bm{U}_{t_1})\T\varphi(\bm{U}_{t_2})\Biggr)^{1/2}\!\!\!\!.
\end{align*}
If we can choose $\varphi(\cdot)$ to be a kind of ``distributional embedding'', for example, in the sense that the two statistics --- $(1/\tau)\sum_{t_1=1}^{\tau} \varphi(\hat{\bm{U}}_{t_1})$ and $(1/\ngen)\sum_{t_2=1}^{\ngen} \varphi(\bm{U}_{t_2})$ --- contain all empirical moments of $\hat{U}$ and $U$, respectively, then the MMD criterion can be used as a proxy for measuring whether the two samples have the same distribution.

By \cite{mercer1909},
the inner product $\varphi(\hat{\bm{u}}_t)\T\varphi(\bm{u}_t)$ can be computed in a reproducing kernel Hilbert space by $K(\hat{\bm{u}}_t,\bm{u}_t)$, where $K(\cdot,\cdot): \IR^{d^{*}} \times \IR^{d^{*}} \mapsto \IR$ denotes a kernel
similarity function. Hence, for a given kernel function $K(\cdot,\cdot)$, the $\MMD$ statistic above is equivalent to
\begin{align}
\MMD(\hat{U},U;K)=\Biggl(\frac{1}{\tau^2} \sum_{t_1=1}^{\tau}\sum_{t_2=1}^{\tau}\!\!K(\hat{\bm{U}}_{t_1},\hat{\bm{U}}_{t_2})- \!\frac{2}{\tau\ngen} \sum_{t_1=1}^{\tau}\sum_{t_2=1}^{\ngen}\!\!K(\hat{\bm{U}}_{t_1},\bm{U}_{t_2}) +\!\frac{1}{\ngen^2} \sum_{t_1=1}^{\ngen}\sum_{t_2=1}^{\ngen}\!\! K(\bm{U}_{t_1},\bm{U}_{t_2})\Biggr)^{1/2}\!\!\!\!.\label{def:MMD}
\end{align}
If $K(\cdot,\cdot)$ is chosen to be a so-called universal kernel function, such as a Gaussian or Laplace kernel, then the associated implicit embedding $\varphi: \mathbb{R}^{d^*} \mapsto \mathbb{R}^{\infty}$ is indeed a ``distributional embedding'' in the sense described above, and one can show that the $\MMD$ converges in probability to $0$ for $\tau,\ngen\to\infty$ if and only if
$\hat{C}_{\text{NN}}=C$ \parencite{gretton2007,gretton2012}.

As suggested by \cite{li2015}, we opt to work with a mixture of Gaussian kernels (rather than a single Gaussian kernel) with different
bandwidth parameters,
\begin{align}
  K(\hat{\bm{u}}_t,\bm{u}_t) =\sum_{i=1}^{\nkrn} K(\hat{\bm{u}}_t,\bm{u}_t;\sigma_i),\label{eq:kernel}
\end{align}
where $\nkrn$ denotes the number of mixture components and
$K(\hat{\bm{u}}_t,\bm{u}_t;\sigma) = \exp(-\lVert
\hat{\bm{u}}_t-\bm{u}_t\rVert_2^2/(2\sigma^2))$ is the Gaussian kernel with
bandwidth parameter $\sigma>0$.

Thus, to train the GMMN $f_{\bm{\theta}}$, we perform the optimization
\begin{align}
\underset{\bm{\theta}}{\min}\  \MMD(\hat{U},(f_{\bm{\theta}}(V);K_{\text{trn}}),\label{eq:optMMD}
\end{align}
where $V=(\bm{V}_1\T,\dots,\bm{V}_{\ngen}\T)\T\in[0,1]^{\ngen\times p}$, the NN transform $f_{\bm{\theta}}$ is understood to be applied row-wise, and $K_{\text{trn}}$ represents the selected mixture of Gaussian kernels used to train the GMMN. The specific choice of the number of mixture components $\nkrn$ and the
bandwidth parameters $\sigma_i$, $i=1,\dots,\nkrn$ that characterize $K_{\text{trn}}$ will be provided in Section~\ref{sec:applications}.

\subsubsection*{Training GMMNs}

We now discuss how we can train the GMMN $f_{\bm{\theta}}$, that is, how we can estimate the parameter vector $\bm{\theta}$. For the sake of convenience, we always simply set $\ngen=\tau$ while training the GMMN. (However, after training we can still generate an arbitrary number of samples from $f_{\hat{\bm{\theta}}}$.)

Directly optimizing  the $\MMD$ loss function in \eqref{def:MMD}, also known as batch optimization, would involve all $\binom{\tau}{2}$ pairs of
observations which is memory-prohibitive even for moderately large $\tau$.
While the Nystr\"{o}m approximation is commonly used to reduce the storage and computational cost of large kernel matrices, it is not a desirable approach for us. This is because, for conventional kernel methods such as support vector machines, the kernel matrix itself is often fixed --- and hence precomputed --- for the corresponding optimization problem, but this is not the case for our optimization problem \eqref{eq:optMMD}.
Instead, we adopt a mini-batch optimization procedure, where we
partition the training dataset into \emph{batches} of size $\nbat$ and use the
batches sequentially to update $\bm{\theta}$. After all the training data are
exhausted, that is, roughly $(\tau/\nbat)$-many gradient steps, one epoch
of the training of the GMMN is completed. Batch optimization results as a special case of this mini-batch optimization procedure when we set $\nbat=\tau$; it can be used with relatively small datasets. To update the parameter vector $\bm{\theta}$, we
utilize the Adam optimizer of \cite{kingma2014b} which uses a ``memory-sticking
gradient'' procedure --- a weighted combination of the current gradient and past
gradients from earlier iterations. The trade-off in utilizing mini-batches, particularly with a smaller batch size $\nbat$, is that it
uses only a partial $\MMD$ loss function when computing each gradient step in the optimization. For a detailed summary of the training procedure, see Algorithm~\ref{algorithm:GMMN:train} in Appendix~\ref{sec:appendix:GMMN}.

\section{Probabilistic forecasts and out-of-sample assessments}\label{sec:probforecast}
We now describe how to make rolling probabilistic forecasts from our estimated MTS model by simulating multiple sample paths forward and constructing \emph{empirical predictive distributions} at each time point. Furthermore, we address how to assess these forecasts with out-of-sample test data. Specifically, we consider a \emph{test period} consisting of time points $\tau+1,\tau+2,\dots,T$.

\subsection{Rolling probabilistic forecasts}

Let $h \in \{1,\dots,T-\tau\}$ denote the simulation horizon. For every $t = \tau,\dots,T-h$,
once all realizations up to and including time $t$ --- namely, $(\bm{X}_s)_{s \leq t}$ --- become available,
we can simulate $\npth$-many $h$-step ahead sample paths conditional on this past information $\mathcal{F}_t=\sigma(\{\bm{X}_s: s \leq t\})$. Let
\begin{align*}
  \hat{X}_{\npth,h|\mathcal{F}_t}=\{\hat{\bm{X}}_{t+1}^{(i)}, \hat{\bm{X}}_{t+2}^{(i)}, \dots, \hat{\bm{X}}_{t+h}^{(i)}\,|\, \mathcal{F}_t\}_{i=1}^{\npth}
\end{align*}
denote such sample paths, encoding an \emph{empirical predictive distribution} at each time point $t+1,t+2,...,t+h$, from which various probabilistic forecasts can be made --- for example, we can forecast
$\mathbb{P}(\bm{X}_{t+2}>x)$ by $(1/\npth)\sum_{i=1}^{\npth} I(\hat{\bm{X}}_{t+2}^{(i)}>x)$.

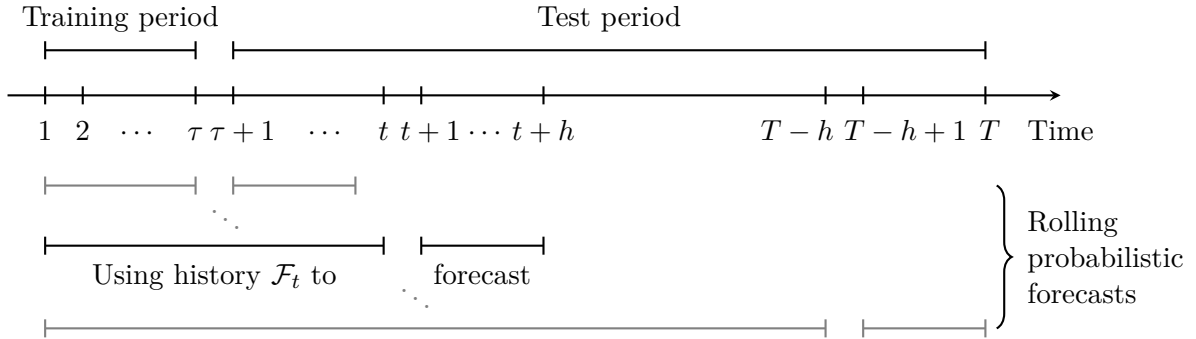
\begin{figure}[htbp]
	\centering
	\begin{tikzpicture}[thick, x=0.5cm]
	\newcommand*{\mylabshift}{0.25}%
	\newcommand*{\myticklength}{1.2mm}%
	\newcommand*{\myy}{6mm}%
	\newcommand*{\myyone}{-12mm}%
	\newcommand*{\myythree}{-20mm}
	\newcommand*{\myyfour}{-31mm}

	\draw[->, >=stealth] (-1,0) -- (27,0) node[anchor=base] at (27,-6mm) {Time};
	\foreach \x in {0,1, 4,5, 9,10, 13.25, 20.75,21.75, 25} \draw(\x, \myticklength) -- (\x, -\myticklength);

	\foreach \x/\what in {0/1, 1/2, 2.5/\cdots, 4/\tau, 5+\mylabshift/\tau+1, 7.5/\cdots, 9/t, 10+\mylabshift/t+1, 11.75/\cdots, 13.25/t+h, 20.75-3*\mylabshift/T-h, 21.75+4.5*\mylabshift/T-h+1, 25.125/T} %
	\node[anchor=base] at (\x,-6mm) {$\what$}; %

	\draw(0,\myy) -- (4,\myy) node[above=1mm, pos=0.5] {Training period};
	\foreach \x in {0,4} \draw(\x, \myy+\myticklength) -- (\x, \myy-\myticklength);
	\draw(5,\myy) -- (25,\myy) node[above=1mm, pos=0.5] {Test period};
	\foreach \x in {5,25} \draw(\x, \myy+\myticklength) -- (\x, \myy-\myticklength);

	\draw[color=black!50] (0,\myyone) -- (4,\myyone);
	\foreach \x in {0,4} \draw[color=black!50] (\x, \myyone+\myticklength) -- (\x, \myyone-\myticklength);
	\draw[color=black!50] (5,\myyone) -- (8.25,\myyone); %
	\foreach \x in {5,8.25} \draw[color=black!50] (\x, \myyone+\myticklength) -- (\x, \myyone-\myticklength);
	\node[color=black!50] at (4.8, -15.5mm) {$\ddots$};
	\draw(0,\myythree) -- (9,\myythree) node[below=1mm, pos=0.5]{Using history $\mathcal{F}_t$ to};
	\foreach \x in {0,9} \draw(\x, \myythree+\myticklength) -- (\x, \myythree-\myticklength);
	\draw(10,\myythree) -- (13.25,\myythree) node[below=1mm, pos=0.5]{forecast};
	\foreach \x in {10,13.25} \draw(\x, \myythree+\myticklength) -- (\x, \myythree-\myticklength);
	\node[color=black!50] at (9.85, -26mm) {$\ddots$};
	\draw[color=black!50] (0,\myyfour) -- (20.75,\myyfour);
	\foreach \x in {0,20.75} \draw[color=black!50] (\x, \myyfour+\myticklength) -- (\x, \myyfour-\myticklength);
	\draw[color=black!50] (21.75,\myyfour) -- (25,\myyfour);
	\foreach \x in {21.75,25} \draw[color=black!50] (\x, \myyfour+\myticklength) -- (\x, \myyfour-\myticklength);

	\draw[thick, decorate, decoration={brace, amplitude=2mm}] (25.3, -12mm) -- +(0,-19mm);
	\node[align=left] at (28.2, -21.5mm) {Rolling\\ probabilistic\\ forecasts};
	\end{tikzpicture}
	\caption{Time periods used for training, testing and the rolling probabilistic forecasts. At each time point $t=\tau,
          \dots,T-h$, the available history $\mathcal{F}_t$ is used to make an $h$-step ahead probabilistic forecast $\hat{X}_{\npth,h|\mathcal{F}_t}$.}\label{fig:timeline}
\end{figure}

Figure~\ref{fig:timeline} provides a schematic illustration of what time periods are used respectively for training and testing, as well as how rolling probabilistic forecasts are made as we move forward step by step in time. In particular, we use the realizations $\bm{X}_1,\dots,\bm{X}_{\tau}$ to train our MTS model and the realizations $\bm{X}_{\tau+1},\dots,\bm{X}_{T}$ to evaluate the rolling probabilistic forecasts made from our trained MTS model. At each particular time point $t = \tau, ..., T-h$, we use all available realizations
up to and including time $t$, that is, $\mathcal{F}_t$, to construct $h$-step ahead probabilistic forecasts $\hat{X}_{\npth,h|\mathcal{F}_t}$, but we do not re-fit the MTS model itself.

  A key component for simulating the paths is the generation of samples from the
estimated dependence model. For fitted parametric copulas
$\hat{C}_{\text{PM}}$, one typically uses a model-specific stochastic
representation to sample $\bm{U}_t$; see, e.g., \cite[Chapter~3]{hofertkojadinovicmaechleryan2018}. For obtaining samples from the standard empirical copula, we can simply resample with replacement from $\hat{\bm{U}}_1,\dots,\hat{\bm{U}}_{\tau}$. For the empirical beta copula, we instead resample with replacement from a collection of uniform order statistics that are indexed by the ranks $R_{t,j}$, $t=1,\dots,\tau$, $j=1,\dots,d^{*}$; see \cite{segers2017} for details. Sampling from
the fitted GMMN $f_{\bm{\hat{\theta}}}$ (with corresponding empirical copula
$\hat{C}_{\text{NN}}$) can be done as follows.
\begin{algorithm}[GMMN sampling]\label{algorithm:GMMN:sampling}
  \begin{enumerate}
  \item Fix the number $\ngen$ of samples to generate from $\hat{C}_{\text{NN}}$.
  \item Draw $\bm{V}_1,\dots,\bm{V}_{\ngen}\isim F_{\bm{V}}$ from the prior distribution.
  \item Return the pseudo-observations of $\bm{U}_s=f_{\hat{\bm{\theta}}}(\bm{V}_s)$, $s=1,\dots,\ngen$.
\end{enumerate}
\end{algorithm}

At any particular time point $t=\tau,\dots,T-h$, we can now utilize Algorithm~\ref{algorithm:GMMN:sampling} along with the fitted marginal time series models in our framework in order to simulate paths $\hat{X}_{\npth,h|\mathcal{F}_t}$ with a fixed horizon $h$, as outlined in Algorithm~\ref{algorithm:forecast}.

\begin{algorithm}[Rolling probabilistic forecasts from a GMMN--GARCH model without re-fitting]\label{algorithm:forecast}
  \begin{enumerate}
  \item Fix the number of sample paths $\npth$ and the simulation horizon $h$.
  \item For $t = \tau,\dots,T-h$ do:
  \begin{enumerate}
  \item\label{algorithm:forecast:step2} Generate  $\bm{U}_{s}^{(i)}$, $i=1,\dots, \npth$, $s=t+1,\dots,t+h$, from
    the fitted GMMN $\hat{C}_{\text{NN}}$ via Algorithm~\ref{algorithm:GMMN:sampling}.
  \item\label{algorithm:forecast:step3} For every $\bm{U}_s^{(i)}$ in Step~\ref{algorithm:forecast:step2}, construct
    $\bm{Y}_{s}^{(i)}=(\hat{F}_{1}^{-1}(U^{(i)}_{s,1}),\dots,\hat{F}^{-1}_{d^{*}}(U^{(i)}_{s,d^{*}}))$.
 	If no dimension reduction
    is utilized, $\hat{F}^{-1}_{j}$, $j=1,\dots,d^{*}$, are the quantile functions of the fitted
    parametric innovation distributions selected as part of the ARMA--GARCH model
    setup; otherwise, they are the empirical quantile functions of $\hat{Y}_{1,j},...,\hat{Y}_{\tau,j}$,
	$j=1,\dots,d^{*}$.
  \item For every $\bm{Y}_s^{(i)}$ in Step~\ref{algorithm:forecast:step3},
  construct samples from the fitted innovation distributions via the transform
    $\bm{Z}_s^{(i)}=\hat{\Upsilon}\bm{Y}_s^{(i)}$; note that $\bm{Y}_s^{(i)} \in \mathbb{R}^{d^*}$ whereas $\bm{Z}_s^{(i)}\in\mathbb{R}^d$.
      \item For each $j=1,\dots,d$, compute $\hat{\sigma}^{2^{(i)}}_{s,j}$, $\hat{\mu}^{(i)}_{s,j}$ and $\hat{X}^{(i)}_{s,j}$, for $i=1,\dots, \npth$ and $s=t+1,\dots,t+h$, via
        \begin{align*}
          &\hat{\mu}^{(i)}_{s,j} =\hat{\mu}_j + \sum_{k=1}^{\hat{p}_{1j}}\hat{\phi}_{jk}(\hat{X}^{(i)}_{s-k,j}-\hat{\mu}_j)+\sum_{l=1}^{\hat{q}_{1j}}\hat{\gamma}_{jl}(\hat{X}^{(i)}_{s-l,j}-\hat{\mu}^{(i)}_{s-l,j}),\\
        &\hat{\sigma}^{2^{(i)}}_{s,j}=\hat{\omega}_j + \sum_{k=1}^{\hat{p}_{2j}} \hat{\alpha}_{jk}(\hat{X}^{(i)}_{s-k,j}-\hat{\mu}^{(i)}_{s-k,j})^2+\sum_{l=1}^{\hat{q}_{2j}} \hat{\beta}_{jl} \hat{\sigma}^{2^{(i)}}_{s-l,j}, \\
        &\hat{X}^{(i)}_{s,j} =
          \hat{\mu}^{(i)}_{s,j} + \hat{\sigma}^{2^{(i)}}_{s,j} Z^{(i)}_{s,j},
      \end{align*}
      where, for $s\leq t$, set $\hat{X}^{(i)}_{s,j}=X_{s,j}$,
      $\hat{\sigma}^{2^{(i)}}_{s,j}=\hat{\sigma}^{2}_{s,j}$, and $\hat{\mu}^{(i)}_{s,j}=\hat{\mu}_{s,j}$ for all $i=1,\dots,\npth$. %
  \item Return $\hat{\bm{X}}^{(i)}_s=(\hat{X}^{(i)}_{s,1},\dots,\hat{X}^{(i)}_{s,d})$, $i=1,\dots,\npth$, $s=t+1,\dots,t+h$.
  \end{enumerate}
  \end{enumerate}
\end{algorithm}

Note that
Step~\ref{algorithm:forecast:step2} in Algorithm~\ref{algorithm:forecast} can be replaced by sampling from a fitted parametric copula $\hat{C}_{\text{PM}}$ or nonparametric copula $\hat{C}_{\text{NPM}}$ to obtain the classical
approach for sampling paths in the copula--GARCH framework.

While Algorithm~\ref{algorithm:forecast} describes how to simulate paths $\hat{X}_{\npth,h|\mathcal{F}_t}$ for any simulation horizon $h$, we will focus on one-step ahead ($h=1$) empirical predictive distributions henceforth.

\subsection{Out-of-sample assessments}\label{sec:assess}
We assess two different aspects of out-of-sample performance with data in the test period; again, see Figure~\ref{fig:timeline}. We are interested in the following questions. First, how well has the cross-sectional dependence structure been captured? Second, how good are the resulting empirical predictive distributions?

\subsubsection{Assessing the quality of the cross-sectional dependence model in the test period}
\label{sec:assess_crossDep}

We can use the $\MMD$ statistic to measure how close the empirical distributions of a fitted GMMN $\hat{C}_{\text{NN}}$, a fitted parametric copula $\hat{C}_{\text{PM}}$ and a nonparametric copula $\hat{C}_{\text{NPM}}$ match the cross-sectional dependence structure of the test dataset, $\bm{X}_{\tau+1},\dots,\bm{X}_{T}$. This cross-sectional dependence structure can be extracted using the fitted (marginal) ARMA--GARCH models and the fitted PCA model (if dimension reduction is applied), as described in the following algorithm.

\begin{algorithm}[Extracting the dependence structure of the test dataset]\label{algorithm:extract}
		\begin{enumerate}
			\item Compute $\hat{\sigma}^2_{t,j}$, $\hat{\mu}_{t,j}$ and $\hat{Z}_{t,j}$ for $t=\tau+1,\dots,T$ and $j=1,\dots,d$ via
                          \begin{align*}
                            &\hat{\mu}_{t,j} =\hat{\mu}_j + \sum_{k=1}^{\hat{p}_{1j}}\hat{\phi}_{jk}(X_{t-k,j}-\hat{\mu}_j)+\sum_{l=1}^{\hat{q}_{1j}}\hat{\gamma}_{jl}(X_{t-l,j}-\hat{\mu}_{t-l,j}),\\
			&\hat{\sigma}^2_{t,j}=\hat{\omega}_j + \sum_{k=1}^{\hat{p}_{2j}} \hat{\alpha}_{jk}(X_{t-k,j}-\hat{\mu}_{t-k,j})^2+\sum_{l=1}^{\hat{q}_{2j}} \hat{\beta}_{jl} \hat{\sigma}^2_{t-l,j},\\
			&\hat{Z}_{t,j} = \frac{X_{t,j}-\hat{\mu}_{t,j}}{\hat{\sigma}_{t,j}}.
			\end{align*}
			\item Obtain a sample from the underlying empirical stationary distribution via the transform $\hat{\bm{Y}}_{t}=\hat{\Upsilon}\T\hat{\bm{Z}}_{t}$, $t=\tau+1,\dots,T$. (Note that $\hat{\bm{Z}}_t\in\mathbb{R}^d$ whereas $\hat{\bm{Y}}_t \in \mathbb{R}^{d^*}$.)
		\item Return the pseudo-observations $\hat{\bm{U}}_t=(\hat{U}_{t,1},\dots,\hat{U}_{t,d^{*}})$ of $\hat{\bm{Y}}_t$, for $t=\tau+1,\dots,T$.
		\end{enumerate}
\end{algorithm}

Let $\hat{U}=(\hat{\bm{U}}\T_{\tau+1},\dots,\hat{\bm{U}}\T_{T})\T \in[0,1]^{(T-\tau) \times d^{*}}$ denote the pseudo-observations obtained from the test dataset via Algorithm~\ref{algorithm:extract}. Furthermore, let $U=(\bm{U}\T_{1},\dots,\bm{U}\T_{\ngen})\T \in[0,1]^{\ngen \times d^{*}}$ denote a sample generated from either $\hat{C}_{\text{NN}}$, $\hat{C}_{\text{PM}}$ or $\hat{C}_{\text{NPM}}$, where we choose $\ngen=T-\tau$ (other choices are possible). We can then compute one realization of the MMD statistic $\MMD(\hat{U},U;K)$ as in~\eqref{def:MMD}. In our analysis in Section~\ref{sec:applications}, we then use an average $\MMD$ statistic based on $\nrep$ repeated samples $U^{(i)}\in[0,1]^{\ngen\times d^{*}},$ $i=1,\dots,\nrep$, given by
\begin{align}
  \AMMD=\frac{1}{\nrep}\sum_{i=1}^{\nrep} \MMD(\hat{U},U^{(i)};K_{\text{tst}}).\label{def:averageMMD}
\end{align}

\begin{remark}\label{rem:AMMD}
  Here, we would like to emphasize that,
  even though the GMMNs are trained to optimize the MMD statistic on the training dataset,
  the AMMD metric defined in~\eqref{def:averageMMD} is still a fair out-of-sample assessment metric since
  \begin{enumerate}
  \item %
    it is applied to compare GMMN-generated samples $U^{(i)}$ against realized innovation copula-samples $\hat{U}$ from the \emph{test dataset}, not the training dataset;
    and
  \item $K_{\text{tst}} \neq K_{\text{trn}}$, so a \emph{different} mixture of Gaussian kernels is used for the assessment than the one used for training the GMMN.
  \end{enumerate}
\end{remark}

\subsubsection{Assessing the quality of empirical predictive distributions}
\label{sec:assess_predDist}

While there exist numerous metrics to assess univariate or multivariate point forecasts, there are only a handful of metrics that can be used to evaluate the quality of dependent multivariate empirical predictive distributions. We now present two such metrics we will use across all numerical examples.

 Firstly, we use a version of the \emph{mean squared error (MSE)} metric defined via the Euclidean norm to assess how well the empirical predictive distribution $\hat{X}_{\npth,1|\mathcal{F}_{t-1}}$
  concentrates around each true value $\bm{X}_t$ in the test dataset, so for $t=\tau+1,\dots,T$. To obtain a single numerical value, we work with an average MSE metric computed over the entire test period $t=\tau+1,\dots,T$, defined by
\begin{align}
  \AMSE=\frac{1}{T-\tau}\sum_{t=\tau+1}^{T} \frac{1}{\npth}\sum^{\npth}_{i=1} \lVert\hat{\bm{X}}^{(i)}_{t}-\bm{X}_t\rVert^2_2.\label{eq:average:MSE}
\end{align}

Secondly, we use the \emph{variogram score} introduced by \cite{scheuerer2015}, which, in our context, assesses if the empirical predictive distribution is biased for the distance between any two component samples. For a single numeric summary, we work with an average variogram score (of order $r$) over the entire test period $t=\tau+1,\dots,T$,
\begin{align}
  \AVS^{r}=\frac{1}{T-\tau} \sum_{t=\tau+1}^{T} \sum_{j_1=1}^{d}\sum_{j_2=1}^{d}\biggl(|X_{t,j_1}-X_{t,j_2}|^r-\frac{1}{\npth}\sum_{i=1}^{\npth} |\hat{X}^{(i)}_{t,j_1}-\hat{X}^{(i)}_{t,j_2}|^r \biggr)^2. \label{eq:average:VS}
\end{align}
As numerically demonstrated by \cite{scheuerer2015}, by focusing on pairwise distances between component samples, this metric discriminates well between various dependence structures.

\section{Applications}\label{sec:applications}

In this section, we demonstrate the flexibility of our GMMN--GARCH approach when
compared to the copula--GARCH approach. To that end, we focus on modeling
multivariate yield curve and exchange-rate time series. Before delving into the
two financial econometrics applications, we will first detail the selection and
setup of component models within our framework that will be utilized for all
examples in this section. Specifically, we will describe the choice of marginal
time series models, the implementation details for GMMN models, and the choice
of parametric and nonparametric copula models used for comparison. All
examples in this section were implemented in \R. GMMN models were fitted with the
package \texttt{gnn}, while various other \R\ packages (see below) were used to fit
ARMA--GARCH models, parametric and nonparametric copula models, and so on. The \R\ packages
\texttt{keras} and \texttt{tensorflow} were used as \R\ interfaces to the
corresponding namesake Python libraries. All GMMN training was carried out on a
single NVIDIA Tesla P100 GPU with 12\,GB RAM; see
  \cite[Appendix~B]{hofertprasadzhu2021} for various aspects on run time measurements.
While we convey our results through plots in this section, we also report them in the form of tables in Appendix~\ref{sec:appendix:app}.

\subsection{Multivariate time series modeling: setup and implementation details}
\subsubsection{Serial dependence models}\label{sec:setup:marginal}
For modeling the marginal time series, we take the broad-brush approach and choose to fit ARMA(1,1)--GARCH(1,1) models with scaled $t$ innovation distributions $F_j(z_j)=t_{\nu_{j}}(z_j\sqrt{\nu_j/(\nu_j-2)})$, $j=1,\dots,d$, to each component sample. As mentioned earlier, these models are popular choices for modeling univariate financial time series. To fit them, we use the function \texttt{fit\_ARMA\_GARCH(, solver = "hybrid")} from the \R\ package \texttt{qrmtools} which relies on \texttt{ugarchfit()} from the \R\ package \texttt{rugarch} (see \cite{ghalanos2019}).

\subsubsection{Cross-sectional dependence models: GMMN architecture and training setup}\label{sec:setup:dependence:GMMN}

In both applications, we experiment with five NN architectures containing a
different number of hidden layers and neurons per layer. For a NN with single
hidden layer, we consider three architectures with 100
($\text{NN}^{1\text{x}}_{100}$), 300 ($\text{NN}^{1\text{x}}_{300}$) and 600
$(\text{NN}^{1\text{x}}_{600})$ neurons per layer, respectively.  While we
generally find that the single-hidden-layer architecture provides sufficient
flexibility for the applications under consideration, we also consider two
deeper NN architectures for the sake of comparison. Due to increased
computational complexity with each added hidden layer, we consider a wide
two-hidden-layer architecture with 600 neurons per layer
($\text{NN}^{2\text{x}}_{600}$) and a narrower three-hidden-layer architecture
with only 300 neurons per layer ($\text{NN}^{3\text{x}}_{300}$). We fix the
activation function in each hidden layer to be ReLU since it offers
computational efficiency via non-expensive and non-vanishing gradients, and the
activation function in the output layer to be sigmoid since our target output
lies in $[0,1]^{d^{*}}$. Additionally, we use batch normalization and dropout
regularization (with a dropout rate of $0.5$) in the hidden layers to facilitate
the training of these five NNs while also controlling for overfitting.

As mentioned earlier in Section~\ref{sec:dependence:GMMN}, we utilize a mixture
of Gaussian kernels $K_{\text{trn}}$ to compute the $\MMD$ statistic defined in~\eqref{def:MMD} when training.
To this end, we fix $\nkrn=6$ and choose bandwidth parameters
$(\sigma_1,\dots,\sigma_6)=(0.001,0.01,0.15,0.25,0.50,0.75)$ as done in \cite{hofertprasadzhu2021}. This hyperparameter setting is
specifically suited for copula samples or pseudo-observations as they lie in
$[0,1]^{d^{*}}$. Furthermore, it was demonstrated in \cite{hofertprasadzhu2021} that
GMMNs trained with this particular specification of the loss function were capable of
learning a wide variety of complex dependence structures.

We choose the dimension of the prior distribution $F_{\bm{V}}$ to be $p=d^{*}$. As a result we obtain a natural $d^{*}$-to-$d^{*}$ GMMN transform $f_{\bm{\theta}}$.
Following common practice, we select
$\bm{V}\sim \N(\bm{0},I_{d^{*}})$, where $I_{d^{*}}$ denotes the identity matrix in
$\IR^{d^{*}\times d^{*}}$. Hence $\bm{V}$ consists of independent standard normal random variables. Since we are working with a modest number of training data points in each of the datasets considered, we opt for a batch optimization procedure presented as a special case ($\nbat=\tau$) of Algorithm~\ref{algorithm:GMMN:train}. For the number of epochs, we choose $\nepo=1000$ which ensures a sufficiently long training period to obtain accurate results.

\subsubsection{Cross-sectional dependence models: parametric copulas}\label{sec:setup:dependence:copula}

For a comparison with GMMN--GARCH models, we also present results for a number
of different parametric copula models $C_{\text{PM}}$. These include Gumbel
copulas, normal copulas with exchangeable correlation matrices, $t$ copulas with
exchangeable and with unstructured correlation matrices and vine copulas. For
all copulas except vines, we use maximum pseudo-likelihood estimation via the
function \texttt{fitCopula(, method = "mpl")} from the \R\ package
\texttt{copula}. We can then generate samples from the fitted copulas via
\texttt{rCopula()}. For vine copulas, we use the \texttt{RvineStructureSelect()}
function from the \R\ package \texttt{VineCopula} to fit a regular-vine
(R-vine) copula, where the tree structure is selected using Dissmann's
algorithm \parencite{dissmann2013} and the pair-copula families are selected
using the AIC criteria. All parametric pair-copula families implemented in
the \R\ package \texttt{VineCopula} are considered when fitting the R-vine
copula. We also produce results for the independence copula which serves as a
simple benchmark model.

\subsubsection{Cross-sectional dependence models: nonparametric copulas}\label{sec:setup:dependence:npcopula}
Additionally, we compare GMMN--GARCH models with certain nonparametric copula
models. The latter include the standard empirical copula and the smoothed
empirical beta copula estimators. We use the $\texttt{empCopula()}$ and
$\texttt{rCopula()}$ functions from the \R\ package \texttt{copula} to fit and
simulate from these two types of empirical copulas.

\subsection{Yield curve modeling}

Analyzing and modeling \emph{zero-coupon bond (ZCB) yield curves}, also referred to as the \emph{term structure of interest rates}, is a critical task in various financial and economic applications. While early research in this area is often  solely focused on constructing models of yield curves based on economic theory, the seminal work by \cite{diebold2006} focused on the critical task of yield curve forecasting.

The primary approach showcased in \cite{diebold2006} was the embedding of autoregressive models within the parametric structure of the three factor Nelson--Siegel model \parencite{nelson1987} which intuitively characterizes the level, slope and curvature of the yield curve. Since then various approaches for forecasting yield curves have been investigated; see \cite{diebold2013} for an overview and \cite{caldeira2016} for a recently proposed forecast combination approach. Most models proposed and reviewed in the literature are particularly designed towards constructing point forecasts for yield curves. Such point forecasts are typically useful in bond portfolio optimization and in the pricing of certain financial assets. Alternatively, distributional forecasts of ZCB yield curves could potentially be helpful in risk management applications, derivative pricing (via simulation) and economic scenario generation. To that end, in this section, we consider modeling US and Canadian ZCB yield curves using MTS models. We then utilize our fitted GMMN--GARCH models to obtain empirical predictive distributions of these ZCB yield curves.

 \subsubsection{Modeling US and Canadian ZCB data}

 For US treasury ZCB data, we consider a 30-dimensional yield curve constructed from ZCBs with times to maturity ranging from 1 to 30 years in annual increments. For Canadian ZCB data, we consider a 120-dimensional yield curve constructed from ZCBs with times to maturity ranging from 0.25 to 30 years in quarterly increments. Refer to the \R\ package \texttt{qrmdata} for further details about these data. In particular, we consider these multivariate time series in the time period from 1995-01-01 to 2015-12-31 (2015-08-31 for the Canadian data), treating data from 1995-01-01 to 2014-12-31 as the training set and the remainder as the test dataset.

As a pre-processing step, we begin by applying a simple difference transform to the original time series. We then take the transformed series to be the series $\bm{X}_t$ that we work with.

Following our framework, we first model the marginal time series using the ARMA--GARCH model setup described in Section~\ref{sec:setup:marginal} with $\mu_j=0$, $j=1,\dots,d$. Since these data are relatively high-dimensional ($d=30$ for the US data and $d=120$ for the Canadian data), we apply PCA to the standardized residuals $\hat{\bm{Z}}_t$ for dimension reduction. Yield curves are indeed amenable to good approximations via lower dimensional representations; various dimension reduction techniques such as factor models have been incorporated by various yield curve models~\parencite[see, e.g.,][]{diebold2006}.
We choose the number of top principal components $k$ to construct the lower dimensional representation for each dataset as follows. We select the smallest $k\geq 3$ such that the first $k$ principal components account for at least 95\% of the total variance in the standardized residuals $\hat{\bm{Z}}_t$. For the US data, this choice is $k=3$; for the Canadian data, it is $k=4$.

\subsubsection{Assessment}\label{sec:applications:assess:US:CA}

We evaluate the performance of our models on the test dataset using the metrics discussed in Section~\ref{sec:assess}.
First, we compute the $\AMMD$ metric~\eqref{def:averageMMD} using $\nrep=100$ replications to assess the quality of the dependence models in the test period; see the explanations in Remark~\ref{rem:AMMD} for why the $\AMMD$ metric is a fair out-of-sample assessment metric.  For the Gaussian mixture kernel $K_{\text{tst}}$ in~\eqref{def:averageMMD}, we fix $\nkrn=5$ and select the bandwidth parameters $\bm{\sigma}=(0.1,0.3,0.5,0.7,0.9)$. Then, to assess if capturing the underlying cross-sectional dependence structure well translates to better one-day-ahead empirical predictive distributions, we compute the $\AMSE$ metric~\eqref{eq:average:MSE} and the $\AVS^{r}$ metric~\eqref{eq:average:VS} using $\npth=1000$ simulated paths. For the average variogram score metric $\AVS^{r}$, a typical choice for the order may be $r=0.5$ as stated in \cite{scheuerer2015}. However, since their concluding remarks note that smaller values of $r$ could potentially yield more discriminative metrics when dealing with non-Gaussian data, we choose $r=0.25$.

\begin{figure}[htbp]
   \centering
   \includegraphics[width=0.49\textwidth]{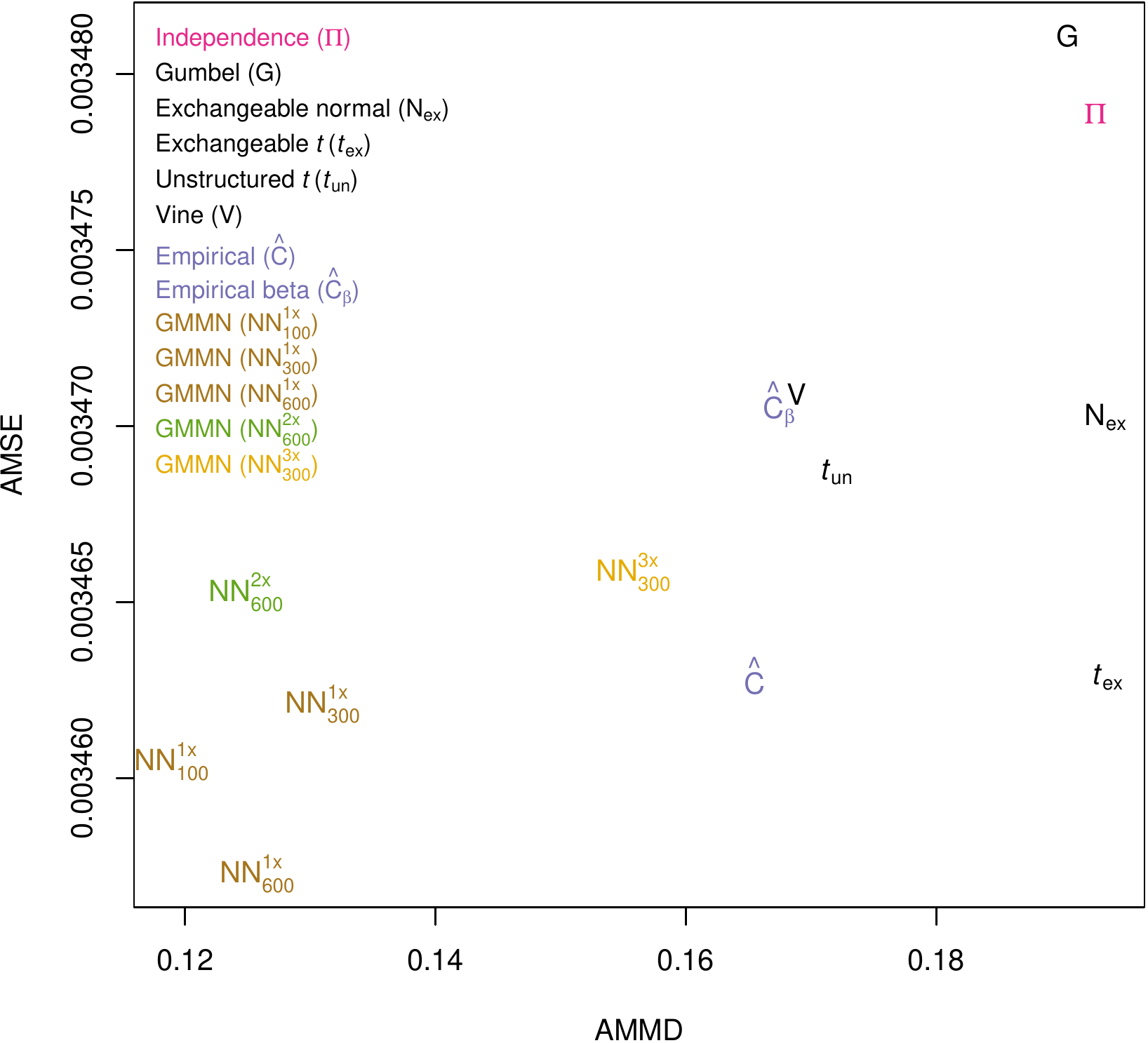}
   \hfill
   \includegraphics[width=0.49\textwidth]{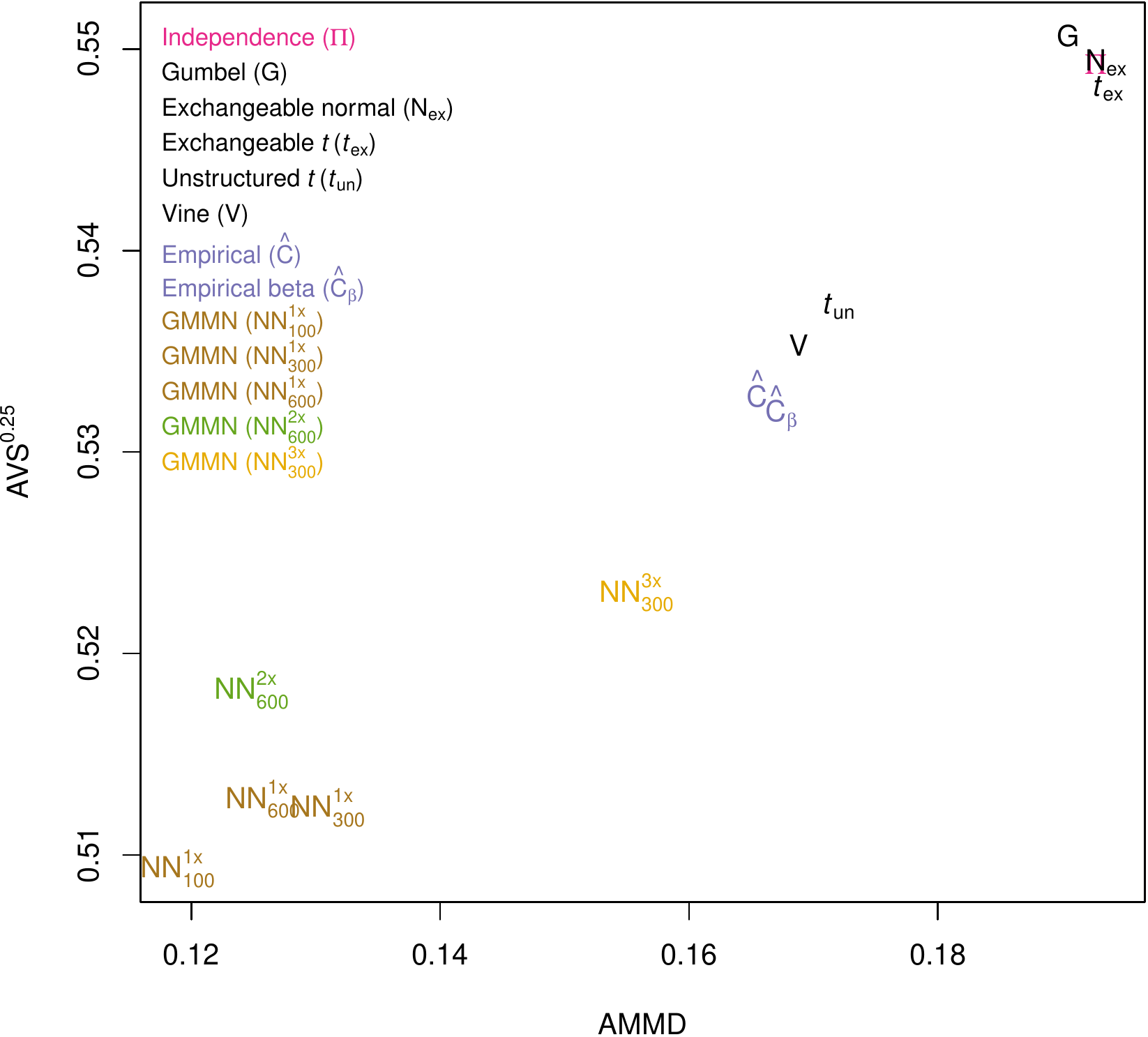}\\[2mm]
   \includegraphics[width=0.49\textwidth]{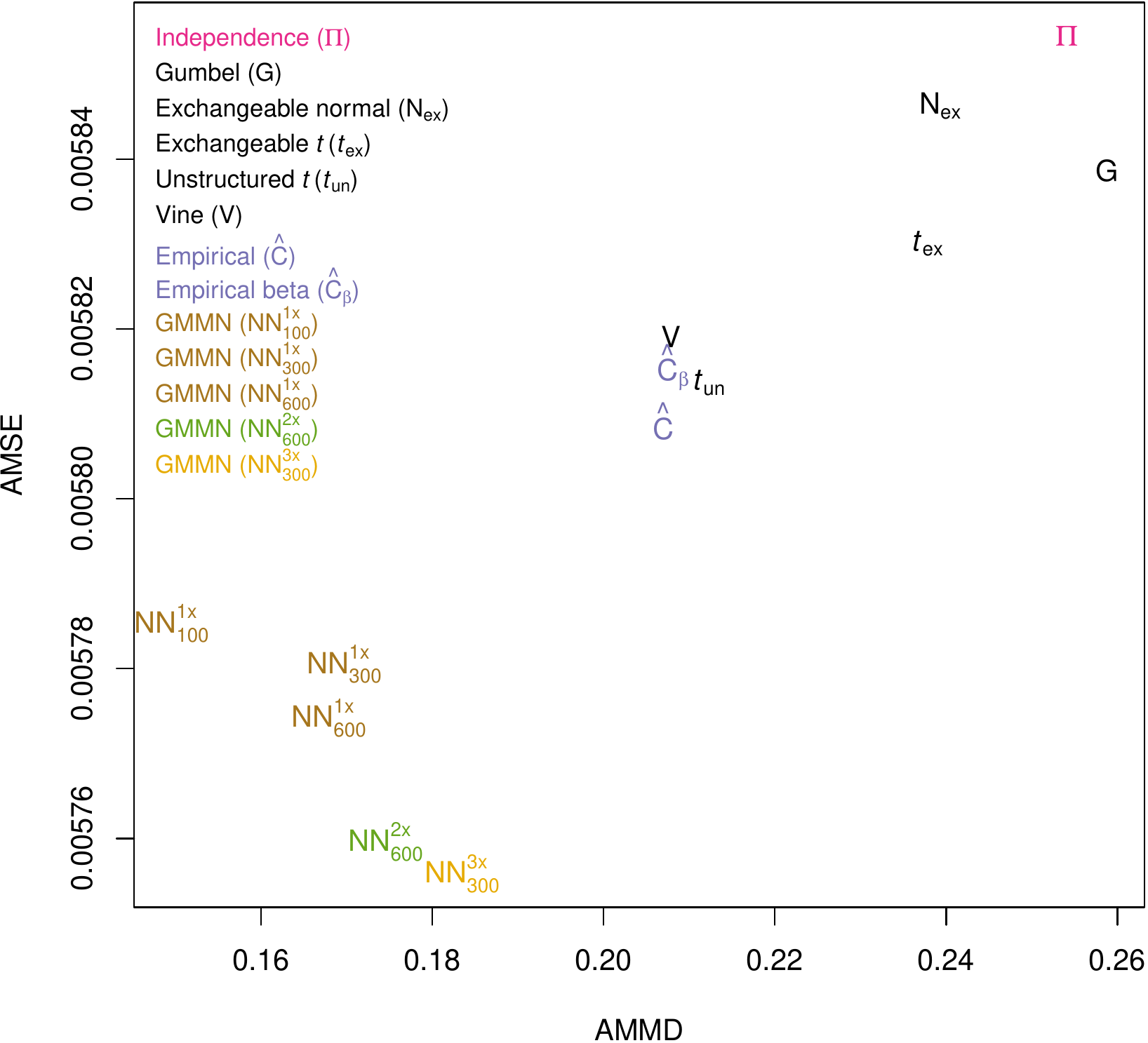}
   \hfill
   \includegraphics[width=0.49\textwidth]{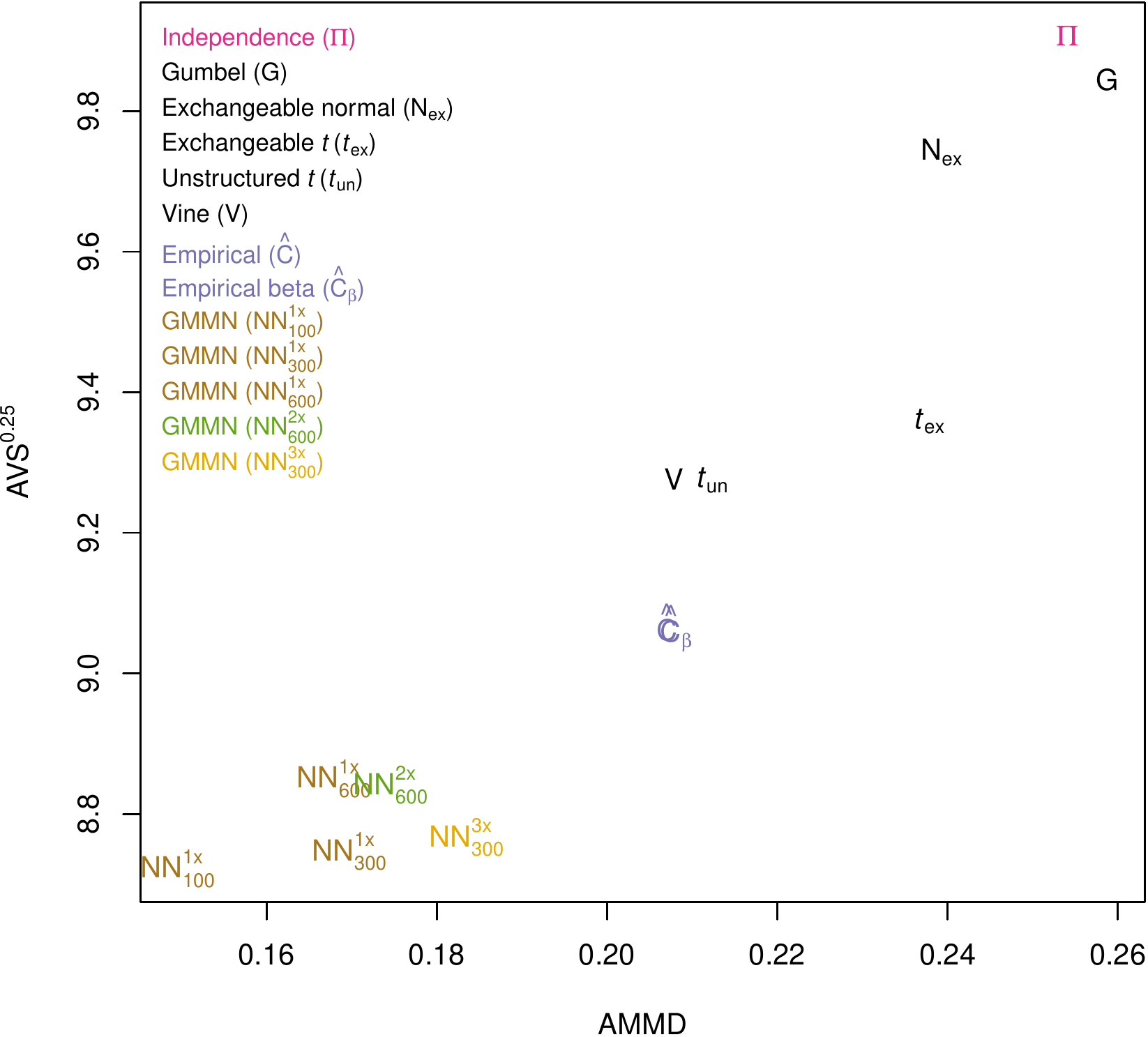}
   \caption{Model assessments for US (top) and Canadian (bottom) ZCB yield curve data. Scatter plots of $\AMSE$ (left) and $\AVS^{0.25}$ (right) computed based on $\npth=1000$ simulated paths versus $\AMMD$ computed based on $\nrep=100$ realizations.  All models incorporate PCA with $k=3$ (US) and $k=4$ (Canadian) principal components.}\label{fig:interest:evals}
 \end{figure}

 Figure~\ref{fig:interest:evals} displays scatter plots of $\AMSE$ (left) and $\AVS^{0.25}$ (right) versus $\AMMD$ for the US (top) and Canadian (bottom) data. For both datasets, samples generated from the five GMMN models (see Section~\ref{sec:setup:dependence:GMMN}) more closely match the underlying cross-sectional dependence structure in their corresponding test datasets than those generated from the five parametric copulas, the independence copula (see Section~\ref{sec:setup:dependence:copula}) and the two nonparametric copulas (see Section~\ref{sec:setup:dependence:npcopula}). Moreover, across the entire spectrum of GMMN--GARCH and copula--GARCH models being studied, it is also clear that better dependence modeling (as measured by the $\AMMD$ metric) does typically translate into better one-day-ahead empirical predictive distributions (as measured by the $\AMSE$ and $\AVS^{0.25}$ metrics). Specifically, almost all GMMN models (with very few exceptions) clearly outperform the best parametric copula model (that is, typically either an R-vine copula or a $t$-copula with unstructured correlation matrix) and the two types of nonparametric copulas in all three metrics --- although among the GMMN models themselves there is not a single best one. Note that more complicated NN architectures ($\text{NN}^{2\text{x}}_{600}$ and $\text{NN}^{3\text{x}}_{300}$) do not necessarily yield better dependence models and (hence) better empirical predictive distributions with respect to the considered metrics.

\subsection{Exchange-rate modeling}

The modeling and analysis of foreign exchange-rate dependence is an important task in risk management applications involving a global portfolio of financial assets. As such, dependent multivariate time series of exchange rates have been previously studied in the copula literature; see, e.g., \cite{patton2006} or \cite{dias2010}. In this section, we consider modeling foreign exchange-rate data with respect to the US dollar (USD) and Pound sterling (GBP) using MTS models. We then utilize our fitted GMMN--GARCH and copula--GARCH models to obtain empirical predictive distributions and  Value-at-Risk (VaR) forecasts for portfolios of exchange-rate assets.

\subsubsection{Modeling USD and GBP exchange-rate data}

For the USD exchange-rate data, we consider the daily exchange rates of Canadian dollar (CAD), Pound sterling (GBP), Euro (EUR), Swiss Franc (CHF) and Japanese yen (JPY) with respect to the USD. For the GBP exchange-rate data, we consider the daily exchange rates of CAD, USD, EUR, CHF, JPY and the Chinese Yuan (CNY) with respect to the GBP. For further details regarding both the USD and GBP exchange-rate data, see the \R\ package \texttt{qrmdata}. In particular, we consider these multivariate time series in the time period from 2000-01-01 to 2015-12-31, treating data up to 2014-12-31 as the training set and the remainder as the test dataset. Due to the fixed peg of the CNY against the USD, particularly prior to August 2005, we do not include it in the USD dataset.

To begin with, we apply the log-returns transformation to the nominal exchange rates and work with the resulting return series for modeling. Following our framework, we start by modeling the marginal time series using the ARMA--GARCH specification as detailed in Section~\ref{sec:setup:marginal}. Since these datasets are relatively low-dimensional ($d=5$ for the USD data and $d=6$ for the GBP data), we do not incorporate any dimension reduction step in this analysis.

\subsubsection{Assessment}

\begin{figure}[htbp]
  \centering
  \includegraphics[width=0.49\textwidth]{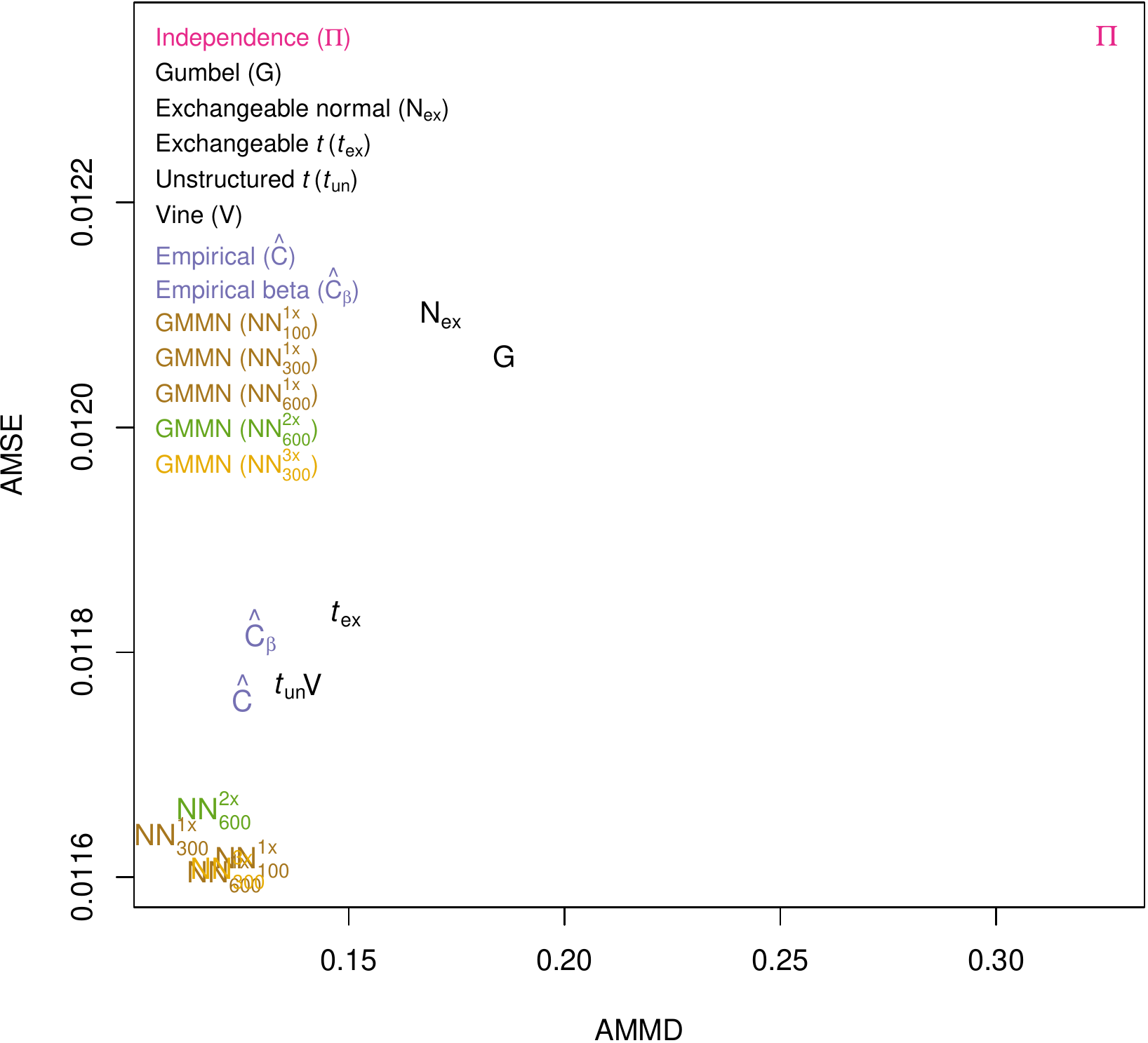}
  \hfill
  \includegraphics[width=0.49\textwidth]{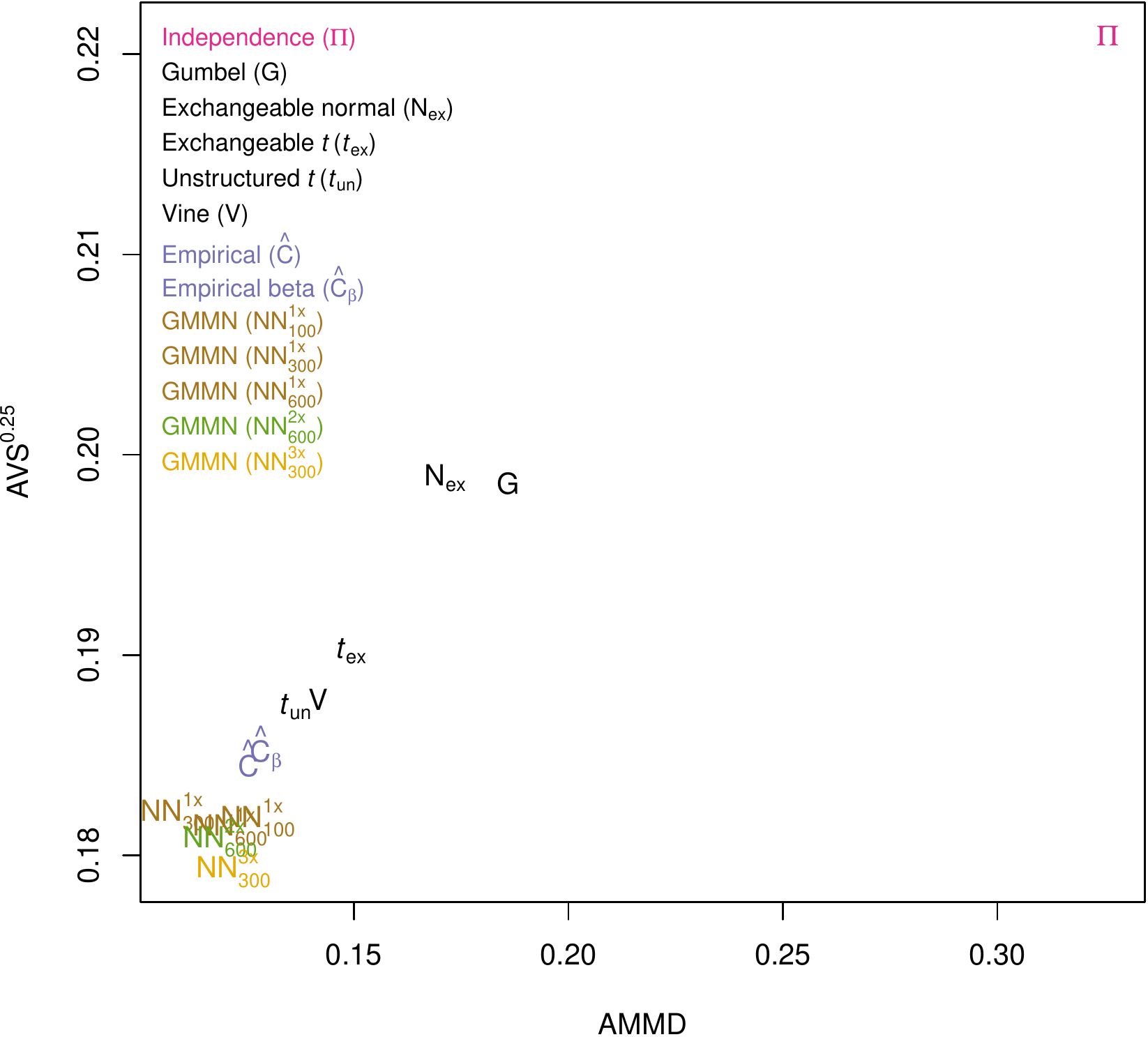}\\[2mm]
  \includegraphics[width=0.49\textwidth]{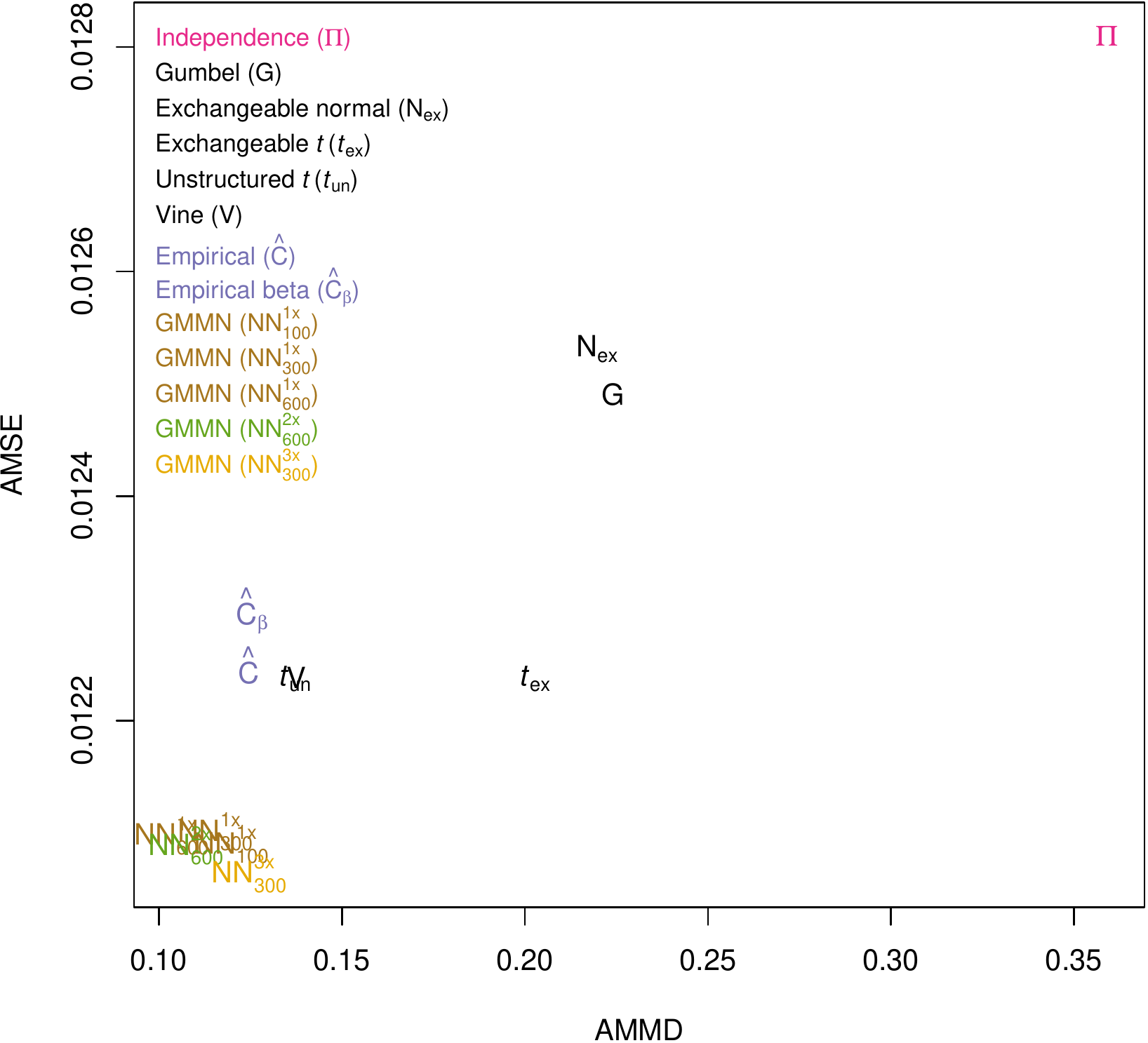}
  \hfill
  \includegraphics[width=0.49\textwidth]{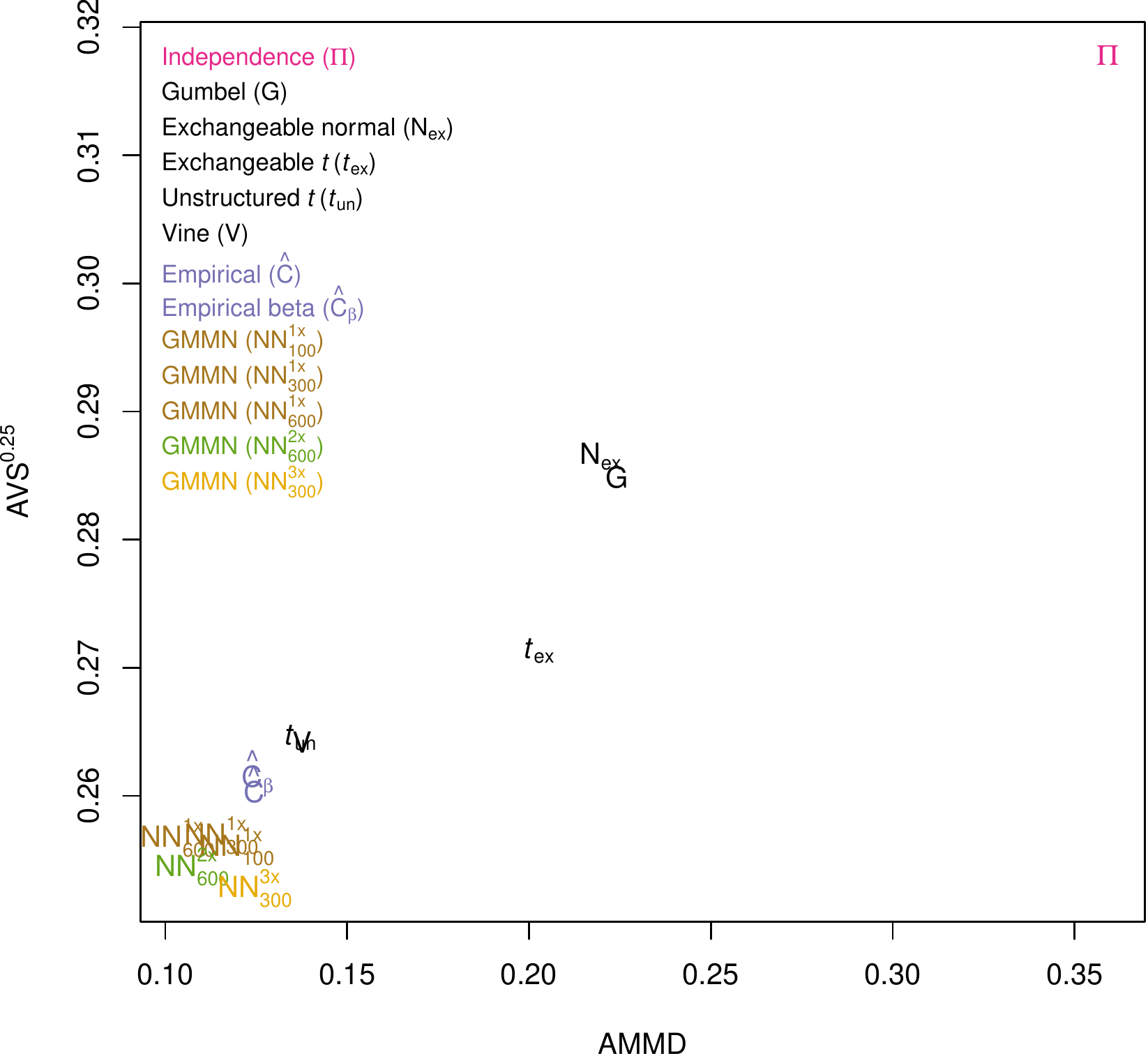}
  \caption{Model assessments for USD (top) and GBP (bottom) exchange-rate data. Scatter plots of $\AMSE$ (left) and $\AVS^{0.25}$ (right) computed based on $\npth=1000$ simulated paths versus $\AMMD$ computed based on $\nrep=100$ realizations.}\label{fig:exchange:evals}
\end{figure}

Following the setup in Section~\ref{sec:applications:assess:US:CA}, we evaluate the performance of our models with the $\AMMD$, $\AMSE$ and $\AVS^{0.25}$ metrics on the test dataset. Figure~\ref{fig:exchange:evals} displays scatter plots of $\AMSE$ (left) and $\AVS^{0.25}$ (right) versus $\AMMD$ for the USD (top) and GBP (bottom) data. We can draw the same conclusions from this figure as those from Figure~\ref{fig:interest:evals}. In addition, here we also observe that the independence copula performs noticeably worse than all other models, whether capturing the dependence structure of the innovation distribution or making probabilistic forecasts.

\subsubsection{Forecasting daily portfolio VaR}

As demonstrated in the previous section, GMMN--GARCH models produce better one-day-ahead empirical predictive distributions when compared with various copula--GARCH models. We can utilize these one-day-ahead empirical predictive distributions to extract forecasts of various quantities of interest in risk management. One such popular quantity is the Value-at-Risk (VaR) of a portfolio.

To begin with, consider the aggregate return $S_t=\sum_{j=1}^{d}X_{t,j}$ at time $t$. Then, the (theoretical) VaR at confidence level $\alpha$ and time $t$ is given by $\text{VaR}_{\alpha}(S_t)=F_{S_t}^{-1}(\alpha)$ where $F_{S_t}^{-1}$ denotes the quantile function of $S_t$. In practice, we can compute the empirical $\alpha$-quantile of $S_t$ from its empirical predictive distribution, $\{\hat{S}^{(i)}_t=\sum_{j=1}^{d}\hat{X}^{(i)}_{t,j}\,|\,\mathcal{F}_{t-1}\}_{i=1}^{\npth}$. We denote the corresponding forecast by $\widehat{\text{VaR}}_{\alpha}(\hat{S}_t)$. Thus, for each MTS model, we compute daily forecasts $\widehat{\text{VaR}}_{\alpha}(\hat{S}_t)$ for every $t=\tau+1,\dots,T$ in the test period. To assess the quality of these forecasts, we can compute the frequency with which $S_t$ actually exceeds the daily forecast $\widehat{\text{VaR}}_{\alpha}(\hat{S}_t)$ over the entire test period. We expect this frequency to be $\alpha$. Hence, we can evaluate our VaR forecasts by measuring the (absolute) error between the actual and the expected exceedance frequency, or simply the \emph{VaR exceedance absolute error}, defined by
\begin{align}
\text{VEAR}_{\alpha}=\bigg| \alpha - \frac{1}{T-\tau}\sum_{t=\tau+1}^{T} \I_{\{S_t<\widehat{\text{VaR}}_{\alpha}(\hat{S}_t)\}}\bigg|.\label{eq:VaR:exceedance:metric}
\end{align}

\begin{figure}[htbp]
  \centering
  \includegraphics[width=0.49\textwidth]{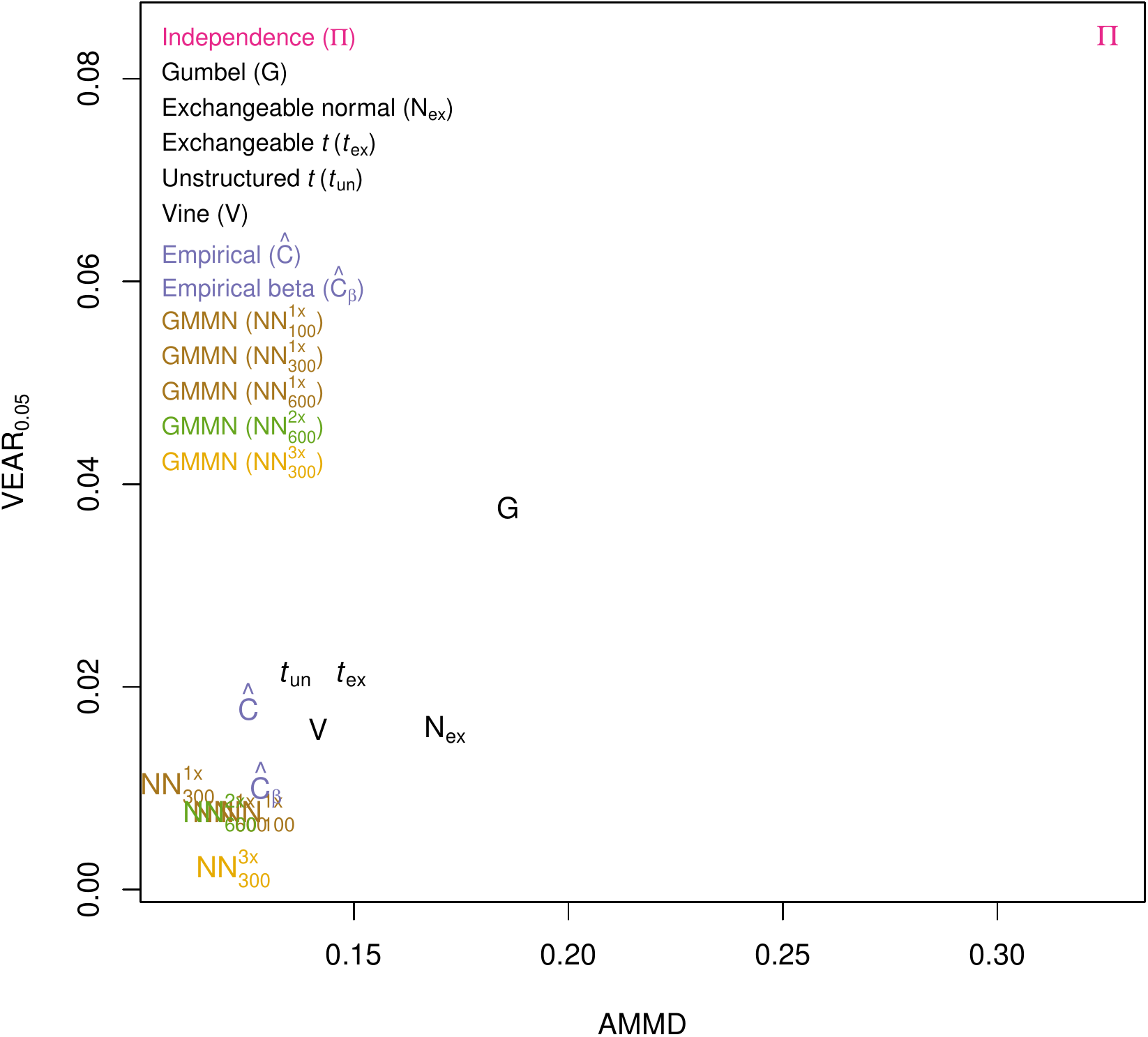}
  \hfill
  \includegraphics[width=0.49\textwidth]{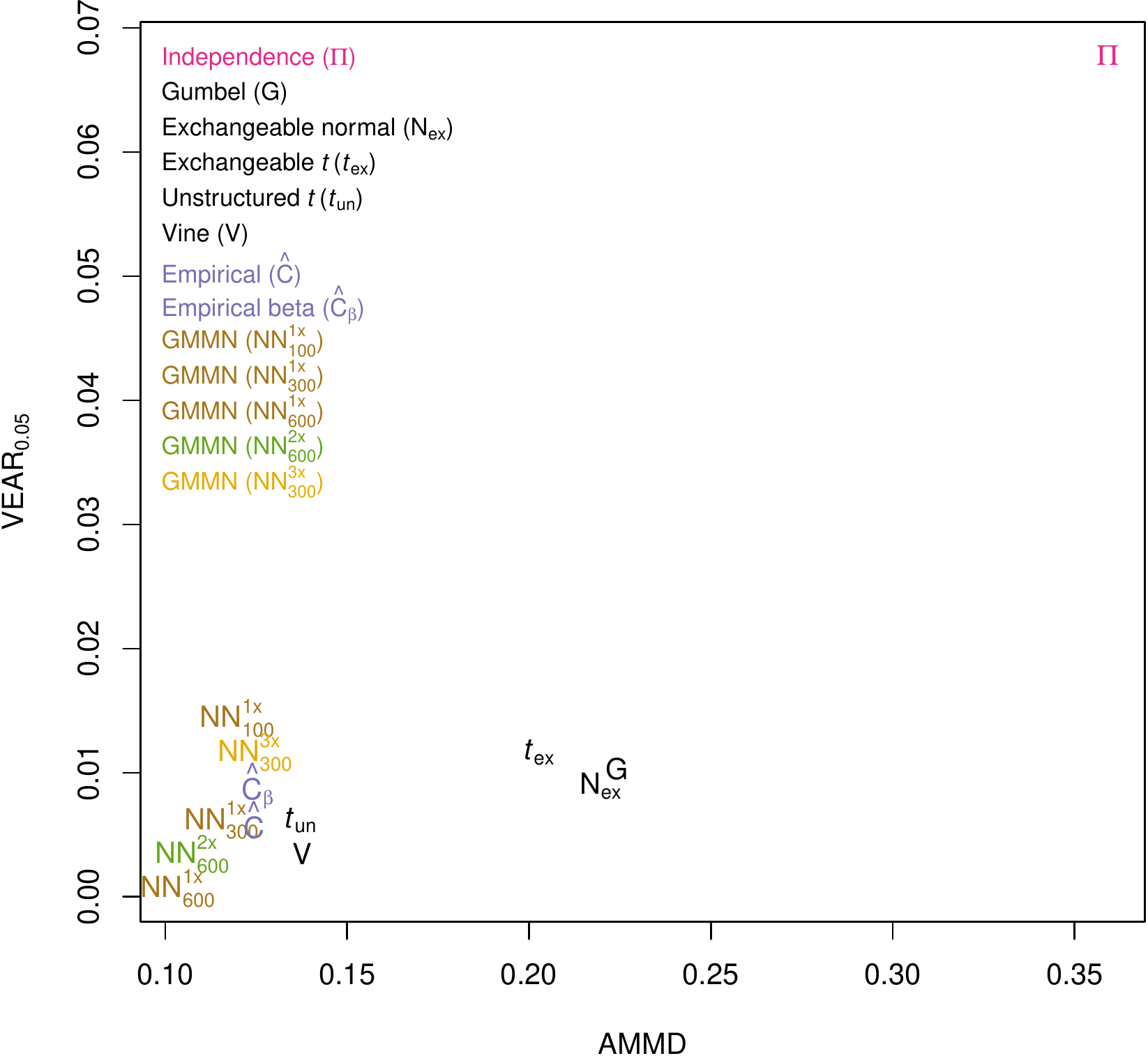}
  \caption{VaR forecast assessments for USD (left) and GBP (right) exchange-rate data. Scatter plots of $\text{VEAR}_{0.05}$ computed based on $\npth=1000$ simulated paths versus $\AMMD$ computed based on $\nrep=100$ realizations.}\label{fig:exchange:VaR}
\end{figure}

Figure~\ref{fig:exchange:VaR} displays scatter plots of $\text{VEAR}_{0.05}$
versus $\AMMD$ for the USD (left) and GBP (right) exchange-rate data. For both
datasets, the GMMN--GARCH models typically produce better daily VaR forecasts of
$\text{VaR}_{0.05}(S_t)$ than the five parametric and two nonparametric
copula--GARCH models do. Particularly, assuming independence among the
exchange-rate returns leads to notably poorer forecasts. However, as expected it
is slightly more difficult to discriminate between the various forecasts when
the evaluation metric is based on a specific $\alpha$-quantile of the empirical
predictive distribution of the aggregated returns as opposed to the entire multivariate
empirical predictive distribution.

\section{Conclusion}\label{sec:concl}

We introduced generative moment matching networks (GMMNs) for modeling the
dependence in MTS data. First,
ARMA--GARCH models are used to marginally model serial dependence. Second, for
high-dimensional MTS data, a dimension reduction method can be
applied. Last, the cross-sectional dependence is modeled by a
GMMN. In the popular copula--GARCH approach, the latter step typically requires us
to find a parametric copula model which fits the given data well. This can already be a
challenging task in moderately large dimensions. By contrast, GMMNs are
highly flexible and easy to simulate from, which is a major advantage
of our GMMN--GARCH approach. The primary objective of fitting these MTS models is to produce empirical predictive distributions, with which we can then forecast various quantities of interest in risk management such as VaR or expected shortfall.

To showcase the flexibility of our GMMN--GARCH framework, we considered modeling ZCB yield curves and foreign exchange-rate returns. Across all the examples considered, we demonstrated that fairly simple GMMNs were able to better capture the underlying cross-sectional dependence than many well-known parametric and nonparametric copulas. Consequentially, we observed that the corresponding GMMN--GARCH models yielded superior one-step ahead empirical predictive distributions. Additionally, for exchange-rate data, we demonstrated that GMMN--GARCH models produced more accurate daily portfolio VaR forecasts as well.

For the first two modeling steps in our framework, we used ARMA--GARCH models
and principal component analysis. However, a variety of other models can be
applied here as long as iid data results as residuals which can then be used to
train GMMNs. A potential avenue for future research involves constructing new and highly flexible MTS models by combining different types of marginal time series models and dimension reduction techniques with GMMNs. In particular, one would be interested in capturing different types of (marginal) temporal dependencies and leveraging more sophisticated dimension reduction techniques for constructing even better higher dimensional time series models.

As \cite{hofertprasadzhu2021} showed, one advantage of GMMNs as dependence models is
that one obtains, for free, a quasi-random number generator from the respective
model. In how far the low discrepancy property propagates to a variance-reduction
effect in forecasted quantities for dependent multivariate time series is also an interesting question
of future research.

\appendix

\section{GMMNs: Additional details}\label{sec:appendix:GMMN}

In this section, we provide some additional details pertaining to the NN architecture and GMMN training.

\subsection*{Feedforward neural networks}
 Let $L$
be the number of hidden layers in the NN and, for each $l=0,\dots,L+1$, let
$d_l$ be the dimension of layer $l$, that is the number of neurons in layer
$l$. Layer $l=0$ refers to the \emph{input layer} which
consists of the input $\bm{v}_t\in \IR^p$ for $d_0=p$, and layer $l=L+1$ refers
to the \emph{output layer} which consists of the output
$\bm{u}_t\in [0,1]^{d^{*}}$ for $d_{L+1}=d^{*}$. The \emph{hidden} layers $l=1,\dots,L+1$ can be
described in terms of the output $\bm{a}_{l-1} \in \IR^{d_{l-1}}$ of layer $l-1$
via
\begin{align*}
\bm{a}_0 &= \bm{v}_t\in\IR^{d_0},\\
\bm{a}_l &= f_l(\bm{a}_{l-1})=\phi_l(W_{l}\bm{a}_{l-1} +\bm{b}_l)\in\IR^{d_l},\quad l=1,\dots,L+1,\\
\bm{u}_t &= \bm{a}_{L+1}\in\IR^{d_{L+1}},
\end{align*}
with \emph{weight matrices} $W_{l} \in \IR^{d_l \times d_{l-1}}$, \emph{bias
	vectors} $\bm{b}_{l} \in \IR^{d_{l}}$ and \emph{activation functions} $\phi_l$; the latter are understood to be
applied componentwise for vector inputs. Some commonly used activation functions include the \emph{sigmoid} activation function $\phi_l(x)=1/(1+e^{x})$ and the \emph{rectified linear unit (ReLU)} activation function $\phi_l(x)=\max\{0,x\}$.
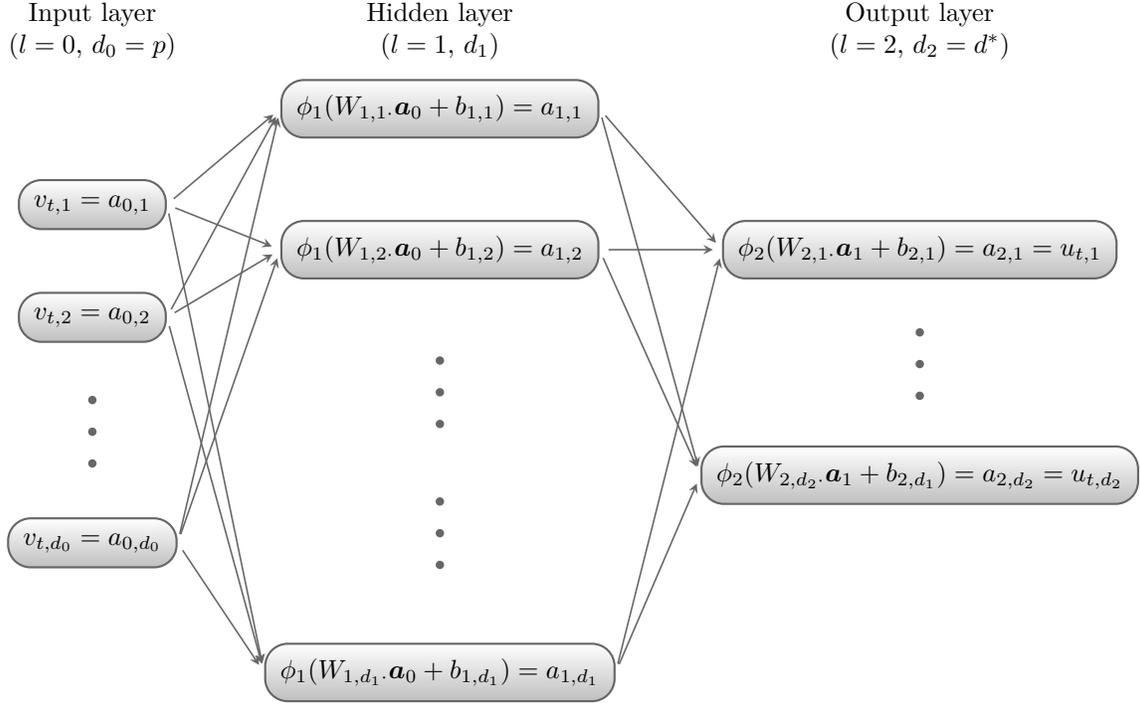
\begin{figure}
	\centering
	\begin{tikzpicture}[
	x = 2.2cm, y = 1.5cm, font = \small,
	neuron/.style={
		rectangle, %
		rounded corners = 3mm,
		thick,
		draw = black!60,
		top color = white,
		bottom color = black!25,
		inner sep = 2mm
	},
	missing/.style={
		color = black!60,
		scale = 3,%
		text height = 0.333cm%
	},
	arrow/.style={
		>= stealth,%
		shorten >=1.2mm,shorten <=1.2mm,
		semithick,
		draw = black!60
	}
	]
	\foreach \j [count=\i] in {1,2,0,3} {%
		\ifnum\j=0%
		\node[missing] at (0, 0.9-\i) {$\vdots$};
		\else
		\node[neuron] (input-\j)%
		at (0, 0.9-\i) {\ifnum\i=4$v_{t,d_0}=a_{0,d_0}$\else$v_{t,\j}=a_{0,\j}$\fi};%
		\fi
	}

	\foreach \j [count=\i] in {1,2,0,0,3} {
		\ifnum\j=0%
		\node[missing] at (2.1, 2-\i*1.25) {$\vdots$};
		\else
		\node[neuron] (hidden-\j)
		at (2.1, 2-\i*1.25) {\ifnum\i=5$\phi_1(W_{1,d_1\cdot}\bm{a}_{0}+b_{1,d_1})=a_{1,d_1}$\else$\phi_1(W_{1,\j\cdot}\bm{a}_{0}+b_{1,\j})=a_{1,\j}$\fi};
		\fi
	}
	\foreach \i in {1,...,3}
	\foreach \j in {1,...,3}
	\draw [arrow, ->] (input-\i.east) -- (hidden-\j.west) node [midway, scale = 0.8] {};

	\foreach \j [count=\i] in {1,0,2} {
		\ifnum\j=0
		\node[missing] at (5, 0.5-\i) {$\vdots$};
		\else
		\node[neuron] (output-\j)
		at (5, 0.5-\i) {\ifnum\i=3$\phi_2(W_{2,d_2\cdot}\bm{a}_{1}+b_{2,d_1})=a_{2,d_2}=u_{t,d_2}$\else$\phi_2(W_{2,\j\cdot}\bm{a}_{1}+b_{2,\j})=a_{2,\j}=u_{t,\j}$\fi};
		\fi
	}
	\foreach \i in {1,...,3}
	\foreach \j in {1,...,2}
	\draw [arrow, ->] (hidden-\i.east) -- (output-\j.west) node [midway, scale = 0.8] {};

	\foreach \l [count=\i from 0] in {{Input layer\\($l=0$, $d_0=p$)}, {Hidden layer\\($l=1$, $d_1$)}, {Output layer\\($l=2$, $d_2=d^{*}$)}} {
		\ifnum\i<2
		\node [align=center, above] at (2.1*\i, 1.1) {\l};
		\else
		\node [align=center, above] at ({5*(\i-1)}, 1.1) {\l};
		\fi
	}
	\end{tikzpicture}
	\caption{Structure of a NN with input $\bm{v}_t=(v_{t,1},\dots,v_{t,d_0})$, $L=1$ hidden layer with output $\bm{a}_1=f_1(\bm{a}_0)=\phi_1(W_1\bm{a}_0+\bm{b}_1)$ and output layer with output $\bm{u}_t=\bm{a}_2=f_2(\bm{a}_1)=\phi_2(W_2\bm{a}_1+\bm{b}_2)$; note that, $W_{l,j\cdot}$ denotes the $j$th row of $W_l$ and $b_{l,j}$ the $j$th element of $\bm{b}_l$.}
	\label{fig:NN}
\end{figure}
The NN $f_{\bm{\theta}}: \IR^p \leftarrow [0,1]^{d^{*}}$ can then be written as the composition
\begin{align*}
f_{\bm{\theta}}=f_{L+1} \circ f_L\circ \dots \circ f_2 \circ f_1,
\end{align*}
with its flattened parameter vector
$\bm{\theta}=(W_{1},\dots,W_{L+1},\bm{b}_{1},\dots,\bm{b}_{L+1})$. Figure~\ref{fig:NN} visualizes this construction and the notation we use. GMMNs are such type of NNs which, for training (that is, the fitting of $\bm{\theta}$), utilize a specific loss function introduced next.

\subsection*{GMMN training}
Algorithm~\ref{algorithm:GMMN:train} describes in detail the training of the GMMN
$f_{\bm{\theta}}$ using a mini-batch optimization procedure.

\begin{algorithm}[Training GMMNs]\label{algorithm:GMMN:train}
	\begin{enumerate}
		\item Fix the number $\nepo$ of
		epochs %
		and the sample size per batch (the so-called \emph{batch size})
		$1\le\nbat\le\tau$, where $\nbat$ is assumed to divide $\tau$. Initialize
		the epoch counter $k=0$ and the GMMN's parameter vector
		$\bm{\theta}=\bm{\theta}^{(0)}$; we follow \cite{glorot2010} and initialize
		the components of $\bm{\theta}^{(0)}$ as
		$W_l\sim \U(-\sqrt{6/(d_l+d_{l-1})},\sqrt{6/(d_l+d_{l-1})})^{d_l \times
			d_{l-1}}$ and $\bm{b}_{l}=\bm{0}$ for $l=1,\dots,L+1$.
		\item Initialize the vectors $\bm{m}^{(0)}_1=\bm{0}$ and $\bm{m}^{(0)}_2=\bm{0}$, where $\bm{m}^{(0)}_1$ and $\bm{m}^{(0)}_2$ have the same dimension as the parameter vector $\bm{\theta}$. Following \cite{kingma2014b}, we fix the exponential decay rates $\beta_1=0.9$ and $\beta_2=0.999$, the step size $\alpha=0.001$ and the smoothing constant $\varepsilon=10^{-8}$.
		\item For epoch $k=1,\dots,\nepo$, do:
		\begin{enumerate}
			\item Randomly partition the training sample
			$\hat{\bm{U}}_1,\dots,\hat{\bm{U}}_{\tau}$ and the prior distribution
			sample $\bm{V}_1,\dots,\bm{V}_{\tau}$ into corresponding $\tau/\nbat$
			non-overlapping batches
			$\hat{\bm{U}}_1^{(b)},\dots,$ $\hat{\bm{U}}_{\nbat}^{(b)}$ and $\bm{V}_1^{(b)},\dots,$ $\bm{V}_{\nbat}^{(b)}$,
			$b=1,\dots,\tau/\nbat$, of size $\nbat$ each.
			\item For batch $b=1,\dots,\tau/\nbat$, let $r=b+(k-1)\tau/\nbat$ and do:
			\begin{enumerate}
				\item Compute the GMMN output $\bm{U}_s^{(b)}=f_{\bm{\theta}^{(r-1)}}(\bm{V}_s^{(b)})$, $s=1,\dots,\nbat$.
				\item Compute the gradient $\nabla^{(r)}=\frac{\partial}{\partial\bm{\theta}}\MMD(\hat{U}^{(b)},U^{(b)};K_{\text{trn}})$ from the samples $\hat{U}^{(b)}=(\hat{\bm{U}}_1^{(b)\top},\dots,\ \hat{\bm{U}}_{\nbat}^{(b)\top})\T$ and
				$U^{(b)}=(\bm{U}_1^{(b)\top},\dots,\bm{U}_{\nbat}^{(b)\top})\T$ via automatic differentiation.
				\item Update $\bm{m}^{(r)}_1=\beta_1\bm{m}^{(r-1)}_1+(1-\beta_1)\nabla^{(r)}$ and compute the bias corrected version $\tilde{\bm{m}}^{(r)}_1=\bm{m}^{(r)}_1/(1-\beta_1^r)$.
				\item Update $\bm{m}^{(r)}_2=\beta_2\bm{m}^{(r-1)}_2+(1-\beta_2)(\nabla^{(r)})^2$, where all operations are understood to be applied componentwise, and compute the bias corrected version $\tilde{\bm{m}}^{(r)}_2=\bm{m}^{(r)}_2/(1-\beta_2^r)$.
				\item Update the parameter vector $\bm{\theta}^{(r)}=\bm{\theta}^{(r-1)}-\alpha\tilde{\bm{m}}^{(r)}_1/\big(\sqrt{\tilde{\bm{m}}^{(r)}_2}+\varepsilon\big)$, where all operations are understood to be applied componentwise.

			\end{enumerate}
		\end{enumerate}
		\item Return $\hat{\bm{\theta}}=\bm{\theta}^{(\nepo(\tau/\nbat))}$; the fitted GMMN is then $f_{\hat{\bm{\theta}}}$.
	\end{enumerate}
\end{algorithm}

\section{Applications: Tables of empirical results}\label{sec:appendix:app}

In this section, we display tables of empirical results for the applications discussed
in Section~\ref{sec:applications}. For five GMMN models, five parametric copulas, the independence
copula and two nonparametric copulas, Table~\ref{tab:ZCB:metrics} presents their respective $\AMMD$, $\AMSE$ and $\AVS^{0.25}$ metrics
for the US and Canadian ZCB yield curve data, while Table~\ref{tab:FX:metrics} presents their respective $\AMMD$, $\AMSE$, $\AVS^{0.25}$ and $\text{VEAR}_{0.05}$ metrics for the USD and GBP exchange-rate data.

\begin{table}[htbp]
	\centering
	\setlength{\tabcolsep}{12pt}

	\begin{tabular}{c c c c c}
		\toprule
		Data set &\multicolumn{1}{c}{Dependence model} & \multicolumn{1}{c}{$\AMMD$} & \multicolumn{1}{c}{$\AMSE$} & \multicolumn{1}{c}{$\AVS^{0.25}$} \\
		\midrule
	   \multirow{13}{*}{US ZCB}& Independence                  & 0.1927 & \num{3.479e-3} & 0.5492\\
	   &Gumbel                        & 0.1905 & \num{3.481e-3} & 0.5507 \\
	   &Exchangeable normal           & 0.1935 & \num{3.470e-3} & 0.5493  \\
	   &Exchangeable $t$              & 0.1936 & \num{3.463e-3} & 0.5480 \\
	   &Unstructured $t$              & 0.1720 & \num{3.469e-3} & 0.5372 \\
	   &Vine                          & 0.1688 & \num{3.471e-3} & 0.5353 \\
	   &Empirical          & 0.1655 & \num{3.463e-3} & 0.5331 \\
	   &Empirical beta  & 0.1674 & \num{3.471e-3} & 0.5322 \\
		&GMMN ($\text{NN}^{1\text{x}}_{100}$)          & 0.1189 & \num{3.460e-3} & 0.5093 \\
		&GMMN ($\text{NN}^{1\text{x}}_{300}$)          & 0.1310 & \num{3.462e-3} & 0.5123 \\
		&GMMN ($\text{NN}^{1\text{x}}_{600}$)          & 0.1257 & \num{3.457e-3} & 0.5128 \\
		&GMMN ($\text{NN}^{2\text{x}}_{600}$)          & 0.1249 & \num{3.465e-3} & 0.5182 \\
		&GMMN ($\text{NN}^{3\text{x}}_{300}$)          & 0.1558 & \num{3.466e-3} & 0.5230 \\
	\midrule
	\multirow{13}{*}{CAD ZCB} & Independence                  & 0.2541  & \num{5.854e-3}  & 9.907 \\
			   &Gumbel                        & 0.2588 &  \num{5.839e-3} & 9.844 \\
			   &Exchangeable normal           & 0.2393 & \num{5.846e-3}  & 9.741   \\
			   &Exchangeable $t$              & 0.2378 & \num{5.830e-3} & 9.359 \\
			   &Unstructured $t$              & 0.2123 & \num{5.814e-3} & 9.275 \\
			   &Vine                          & 0.2079  & \num{5.819e-3}  & 9.276 \\
			   &Empirical          & 0.2070 & \num{5.809e-3} & 9.072  \\
			   &Empirical beta & 0.2081  & \num{5.815e-3}  & 9.063 \\
			&GMMN ($\text{NN}^{1\text{x}}_{100}$)          & 0.1496 & \num{5.785e-3} & 8.722 \\
			&GMMN ($\text{NN}^{1\text{x}}_{300}$)          & 0.1697 &\num{5.781e-3} & 8.747 \\
			&GMMN ($\text{NN}^{1\text{x}}_{600}$)          & 0.1679 & \num{5.774e-3} & 8.850 \\
			&GMMN ($\text{NN}^{2\text{x}}_{600}$)          & 0.1746 & \num{5.760e-3}  & 8.841 \\
			&GMMN ($\text{NN}^{3\text{x}}_{300}$)          & 0.1835 &  \num{5.756e-3} & 8.767 \\

		\bottomrule
	\end{tabular}
\caption{Model assessment metrics for US (top) and Canadian (bottom) ZCB yield curve data. The $\AMSE$ and $\AVS^{0.25}$ metrics are computed based on $\npth=1000$ simulated paths while the $\AMMD$ metric is computed based on $\nrep=100$ realizations. All models incorporate PCA with $k=3$ (US) and $k=4$ (Canadian) principal components.}\label{tab:ZCB:metrics}
\end{table}

\begin{table}[htbp]
	\centering
	\setlength{\tabcolsep}{12pt}

	\begin{tabular}{c c c c c c}
		\toprule
		Data set&\multicolumn{1}{c}{Dependence model} & \multicolumn{1}{c}{$\AMMD$} & \multicolumn{1}{c}{$\AMSE$} & \multicolumn{1}{c}{$\AVS^{0.25}$}& $\text{VEAR}_{0.05}$ \\
		\midrule
		\multirow{13}{*}{USD FX}&  Independence                  & 0.3257   & \num{1.235e-2}  & 0.2209 & \num{8.425e-2} \\
		&Gumbel                        & 0.1860  &  \num{1.206e-2} &  0.1986 &  \num{3.767e-2} \\
		&Exchangeable normal           & 0.1713  & \num{1.210e-2}  & 0.1988 &  \num{1.575e-2}\\
		&Exchangeable $t$              & 0.1492 & \num{1.183e-2} & 0.1902 & \num{2.123e-2} \\
		&Unstructured $t$              & 0.1363 &
		\num{1.177e-2} & 0.1874 & \num{2.123e-2}  \\
		&Vine                          & 0.1416  &  \num{1.177e-2}  & 0.1878 & \num{1.575e-2}   \\
		&Empirical          & 0.1254  & \num{1.176e-2} & 0.1848 & \num{1.849e-2} \\
		&Empirical beta  & 0.1295  & \num{1.182e-2}  & 0.1853 &  \num{1.027e-2} \\
		&GMMN ($\text{NN}^{1\text{x}}_{100}$)          & 0.1276 & \num{1.162e-2}  & 0.1819 & \num{0.753e-2} \\
		&GMMN ($\text{NN}^{1\text{x}}_{300}$)          & 0.1089 & \num{1.164e-2}  & 0.1822 &  \num{1.027e-2} \\
		&GMMN ($\text{NN}^{1\text{x}}_{600}$)          & 0.1211 & \num{1.160e-2} & 0.1814& \num{0.753e-2} \\
		&GMMN ($\text{NN}^{2\text{x}}_{600}$)          & 0.1188  & \num{1.166e-2}   & 0.1809&  \num{0.753e-2}  \\
		&GMMN ($\text{NN}^{3\text{x}}_{300}$)          & 0.1219  &  \num{1.161e-2} & 0.1793 & \num{0.206e-2} \\
		\midrule
		   \multirow{13}{*}{GBP FX}& Independence                  & 0.3591   & \num{1.281e-2}  & 0.3178  & \num{6.781e-2} \\
		   &Gumbel                        & 0.2241  &  \num{1.249e-2}  &  0.2848  &  \num{1.027e-2} \\
		   &Exchangeable normal           & 0.2197  & \num{1.253e-2}  &  0.2865 &  \num{0.890e-2} \\
		   &Exchangeable $t$              & 0.2026  &  \num{1.224e-2} & 0.2713 &  \num{1.164e-2} \\
		   &Unstructured $t$              & 0.1371 &  \num{1.224e-2}  &  0.2645 & \num{0.616e-2}  \\
		   &Vine                          & 0.1376   &   \num{1.224e-2}   & 0.2642 &  \num{0.343e-2}   \\
		   &Empirical          & 0.1245   & \num{1.225e-2} & 0.2609 &  \num{0.616e-2} \\
		   &Empirical beta & 0.1254  & \num{1.230e-2}  & 0.2617 &  \num{0.890e-2} \\
		&GMMN ($\text{NN}^{1\text{x}}_{100}$)          & 0.1198  & \num{1.209e-2}   & 0.2560 & \num{1.439e-2}  \\
		&GMMN ($\text{NN}^{1\text{x}}_{300}$)          & 0.1154 & \num{1.210e-2}   & 0.2569  &  \num{0.616e-2} \\
		&GMMN ($\text{NN}^{1\text{x}}_{600}$)          & 0.1034 & \num{1.210e-2} & 0.2567 & \num{0.068e-2} \\
		&GMMN ($\text{NN}^{2\text{x}}_{300}$)          &  0.1074   & \num{1.209e-2}   & 0.2544 &   \num{0.343e-2} \\
		&GMMN ($\text{NN}^{3\text{x}}_{600}$)          & 0.1247   &  \num{1.206e-2} & 0.2528 & \num{1.164e-2} \\

		\bottomrule
	\end{tabular}
	\caption{Model assessment metrics for USD (top) and GBP (bottom) exchange-rate data. The $\AMSE$, $\AVS^{0.25}$ and $\text{VEAR}_{0.05}$ metrics are computed based on $\npth=1000$ simulated paths while the $\AMMD$ metric is computed based on $\nrep=100$ realizations. }\label{tab:FX:metrics}
\end{table}

\section{Estimation uncertainty in MTS modeling}\label{sec:appendix:uncertain}
Establishing theoretical results for our MTS model is challenging. In a
time-static setting (i.e., outside the context of MTS), \cite{genest1995}
investigated the maximum pseudo-likelihood estimator (MPLE) of parametric
copulas. In the context of copula--GARCH models, \cite{chan2009} studied the
MPLE of residual parametric copulas ($\hat{C}_{\text{PM}}$ in our framework). In
particular, by constructing a weighted approximation to the empirical
distribution of (marginal) GARCH residuals, they were able to derive asymptotic
normality for the MPLE of $\hat{C}_{\text{PM}}$ under regularity conditions, and
establish that the derived limiting distribution is independent of the
(marginal) GARCH processes.
The dimension-reduction step that we used in some applications makes it harder to prove something similar, even for parametric copulas.
Moving to $\hat{C}_{\text{NN}}$
for our proposed GMMN--GARCH model will require another leap. Even in
the static setting, research on statistical inference for generative models is
fairly sparse. \cite{farrell2021} recently derived non-asymptotic high
probability bounds on the estimation error of deep neural
networks with regression-type loss functions, which, in turn, yielded asymptotic
results for subsequent semi-parametric inferences in certain set-ups. However,
these results are not easily adaptable to generative neural networks with MMD
loss functions.

While some theoretical results in the wider vicinity of the copula--GARCH
approach have been established, albeit under conditions not always easy to
verify in practice, statistical inference for \emph{generative models} in both
static (GMMNs) and dynamic (GMMN--GARCH models) settings remains largely an open
problem.
For example, the cross-sectional dependence models
(Section~\ref{sec:dependence}) --- whether copulas or GMMNs --- are trained from
$\hat{\bm{Y}}_1,\dots,\hat{\bm{Y}}_{\tau}$. The fact that these are estimates
themselves introduces additional uncertainty about the resulting models
$\hat{C}_{\text{PM}}$, $\hat{C}_{\text{NPM}}$ and $\hat{C}_{\text{NN}}$. While
\cite{chan2009} studied the impact on training a parametric copula
$\hat{C}_{\text{PM}}$, it seems much harder to develop similar theories for
training a GMMN $\hat{C}_{\text{NN}}$.

In this appendix, we conduct an empirical study of this impact by bootstrapping
$\hat{\bm{Y}}_1,\dots,\hat{\bm{Y}}_{\tau}$. In principle, we could have
broadened the scope of our investigation by bootstrapping the original data
$\bm{X}_1,\dots,\bm{X}_{\tau}$ instead, which would have allowed us to also
account for the uncertainties introduced in the first two steps
(Sections~\ref{sec:serial} and \ref{sec:pca}). However, cross-sectional
dependence has a much subtler effect on the quality of the final probabilistic
forecasts than serial dependence does. Therefore, once additional uncertainties
about the serial dependence models are considered, it becomes much harder to
distinguish various cross-sectional dependence models from one another, which is
why we prefer a more focused investigation with a more limited scope.

We draw $n_{\text{bt}}$-many bootstrap samples of size $\tau$, that is,
$\hat{\bm{Y}}^{(b)}_1,\dots,\hat{\bm{Y}}^{(b)}_{\tau}$, for
$b=1,\dots,n_{\text{bt}}$. The margins are modeled non-parametrically for each
of the $n_{\text{bt}}$-many bootstrap samples by computing the
pseudo-observations for each sample to obtain
$\hat{\bm{U}}^{(b)}_1,\dots,\hat{\bm{U}}^{(b)}_{\tau}$,
$b=1,\dots,n_{\text{bt}}$. We then fit $n_{\text{bt}}$-many dependence models
(GMMNs or copulas), one for every collection of pseudo-observations. Using these
bootstrapped dependence models together allows us to account for the estimation
uncertainty associated with modeling $\hat{\bm{Y}}_t$. To do so, in
Step~\ref{algorithm:forecast:step2} of Algorithm~\ref{algorithm:forecast}, we
instead generate samples from an equally-weighted mixture of all $n_{\text{bt}}$
dependence models. Additionally, the quantile functions associated with each
mixture component are utilized in Step~\ref{algorithm:forecast:step3} of
Algorithm~\ref{algorithm:forecast}.

\begin{figure}[htbp]
  \centering
  \includegraphics[width=0.32\textwidth]{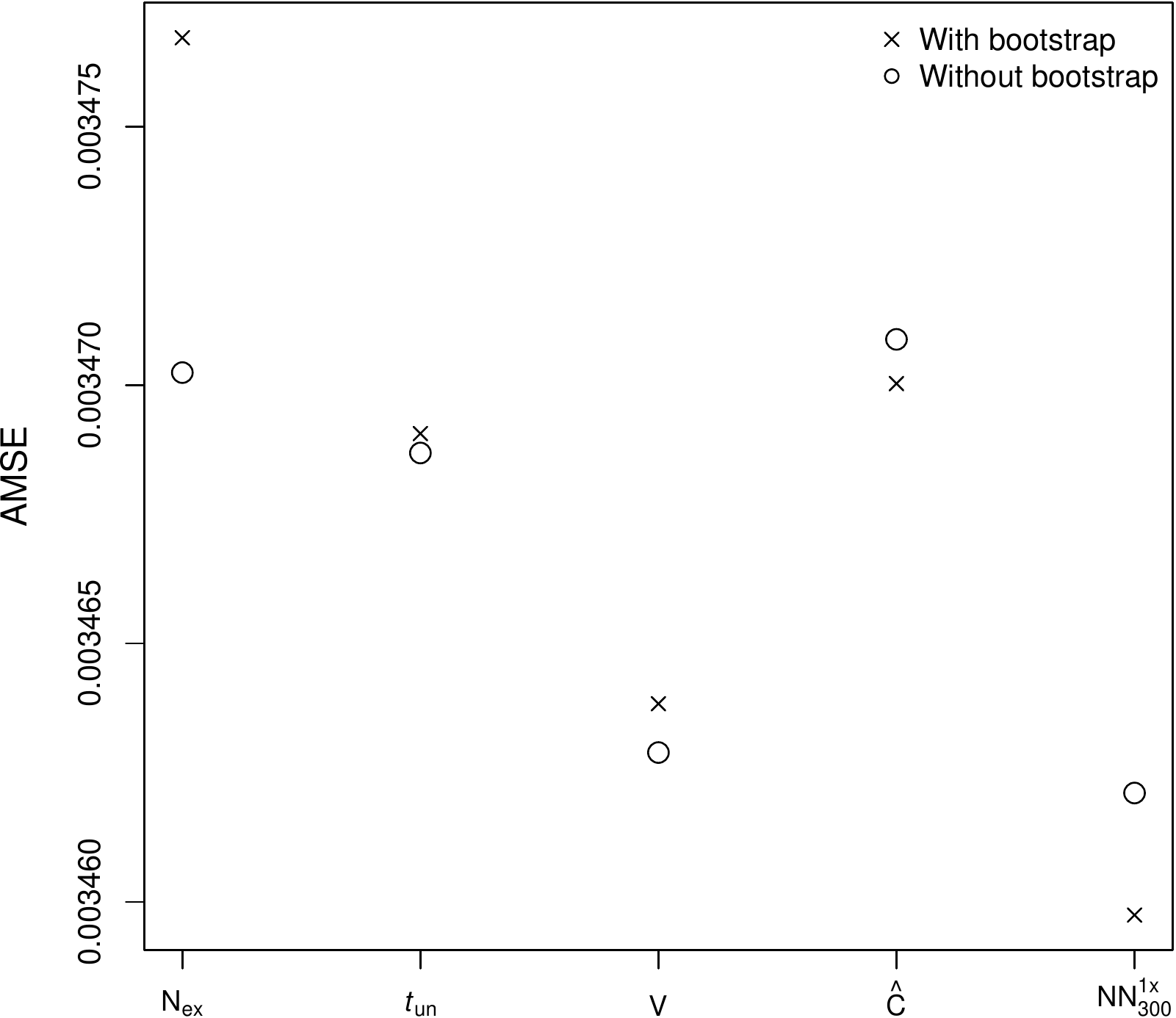}
  \hspace{1.2cm}
  \includegraphics[width=0.32\textwidth]{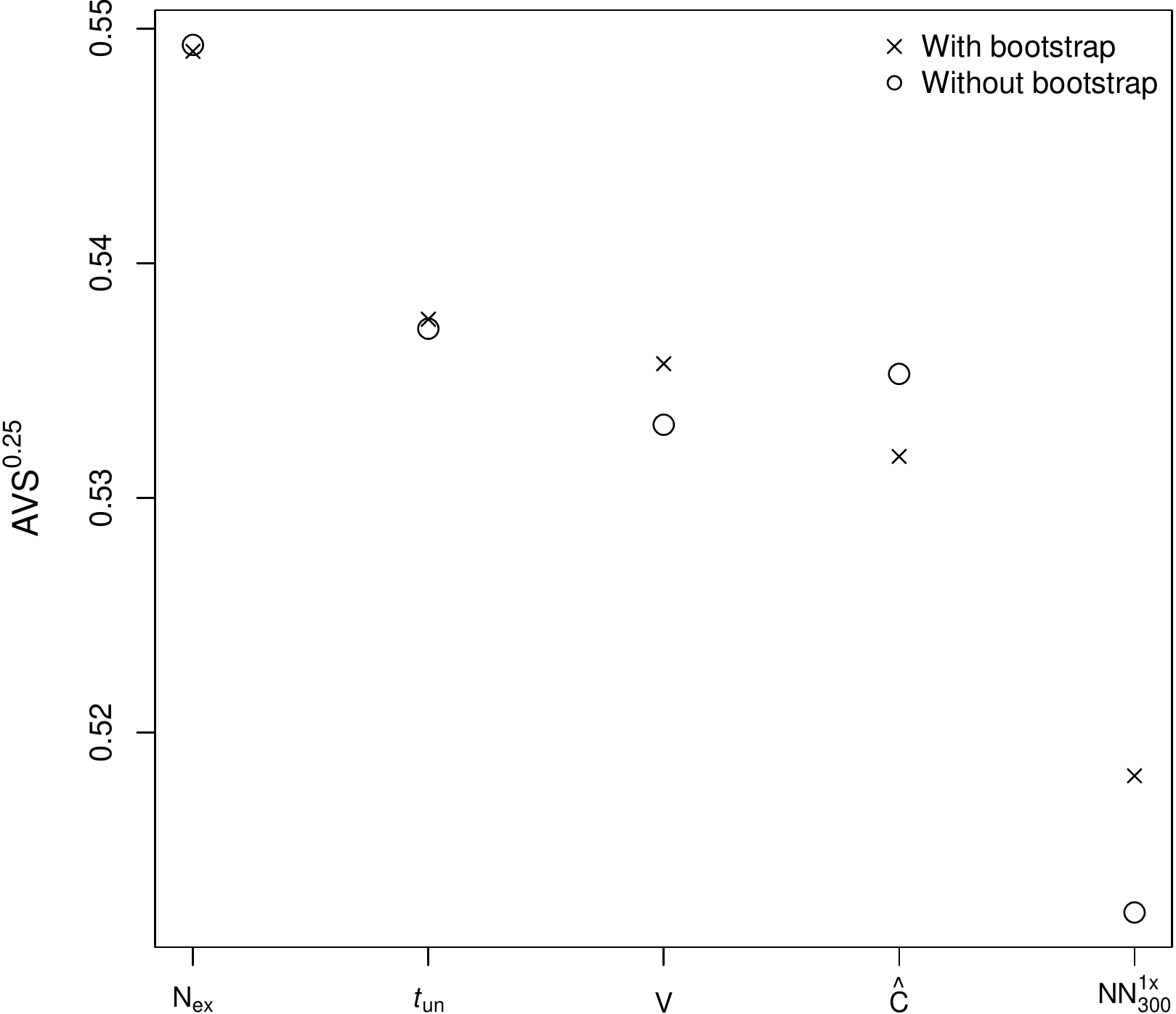}\\[2mm]
  \includegraphics[width=0.32\textwidth]{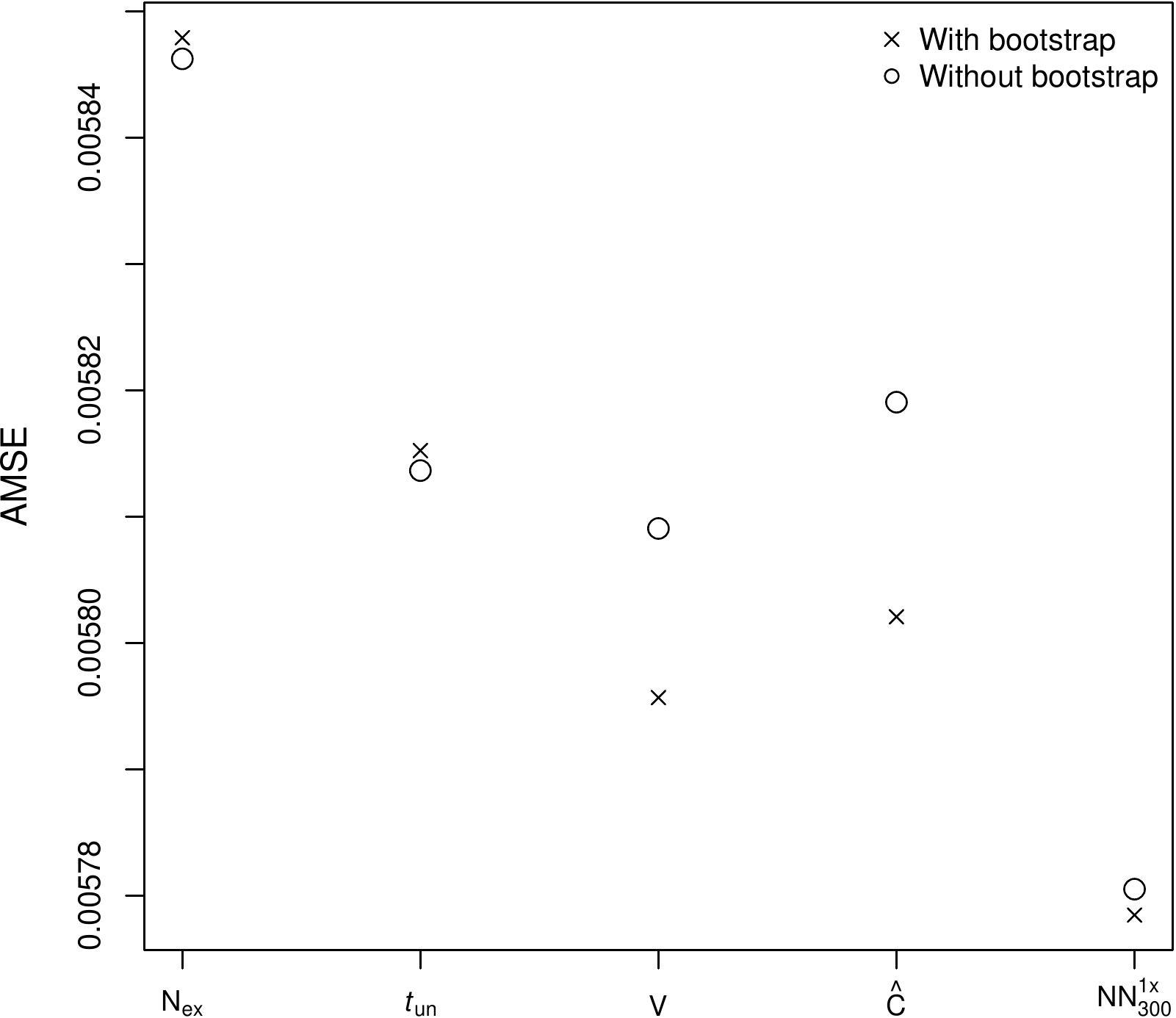}
  \hspace{1.2cm}
  \includegraphics[width=0.32\textwidth]{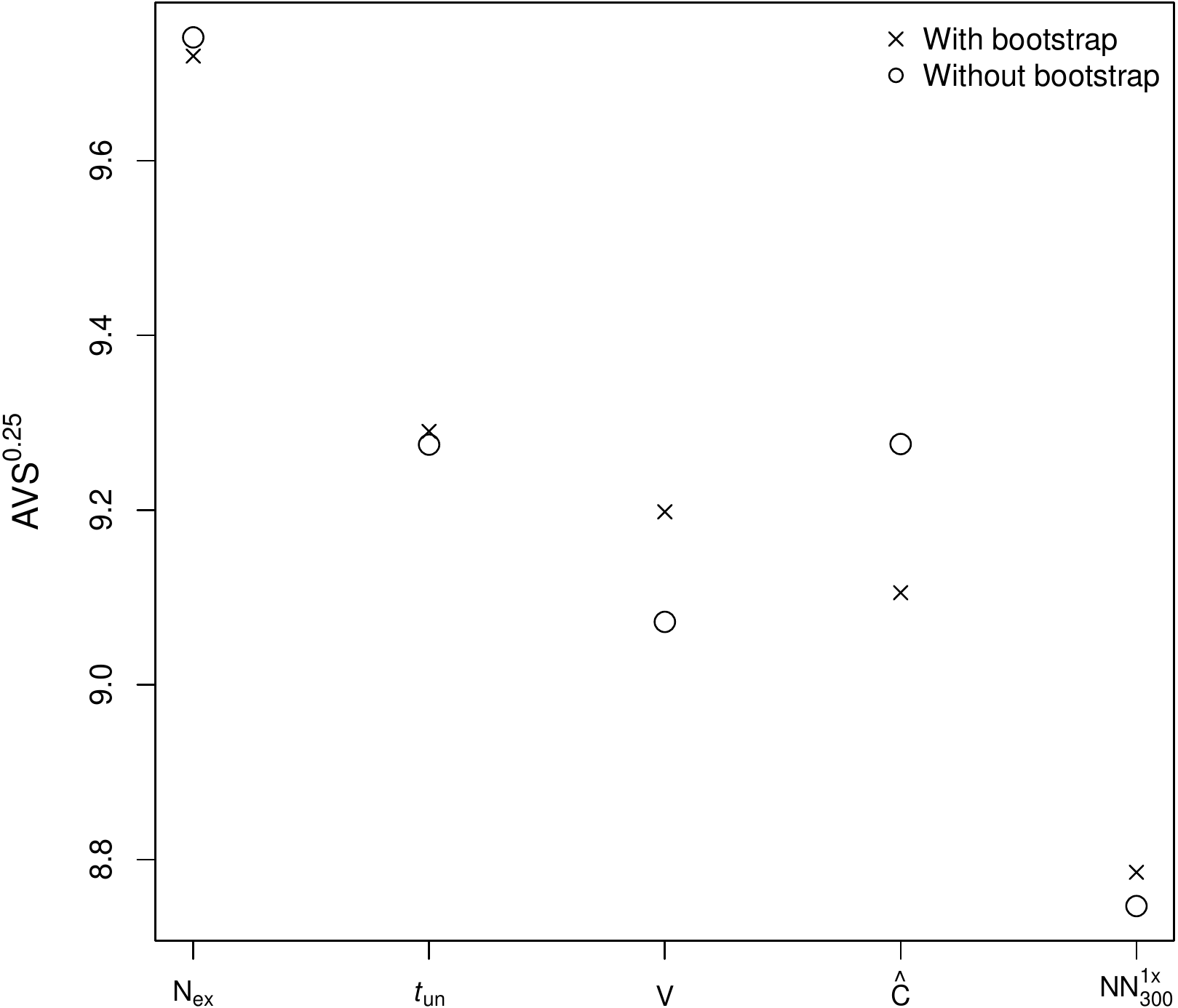}\\[2mm]
  \includegraphics[width=0.32\textwidth]{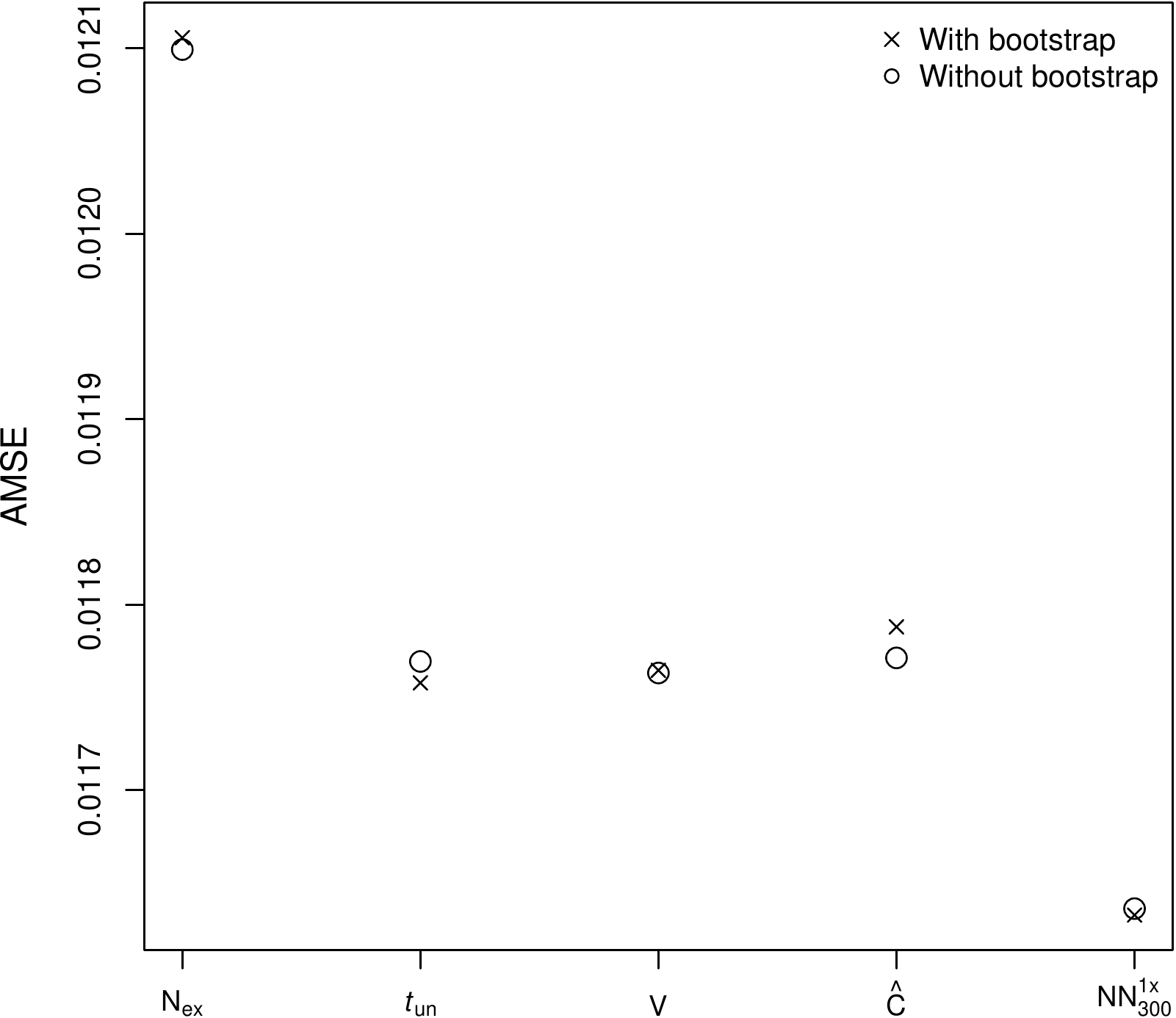}
  \hspace{1.2cm}
  \includegraphics[width=0.32\textwidth]{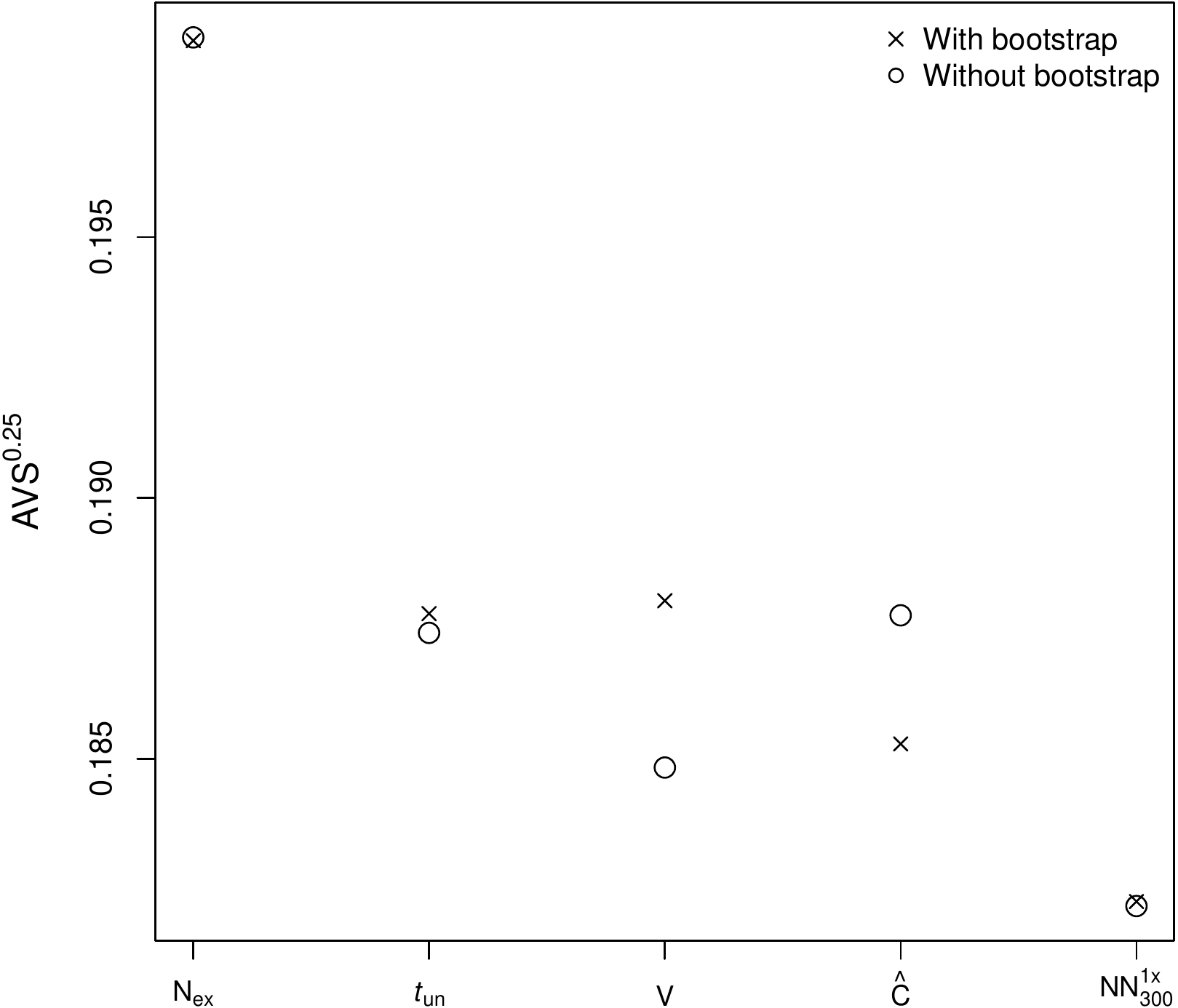}\\[2mm]
  \includegraphics[width=0.32\textwidth]{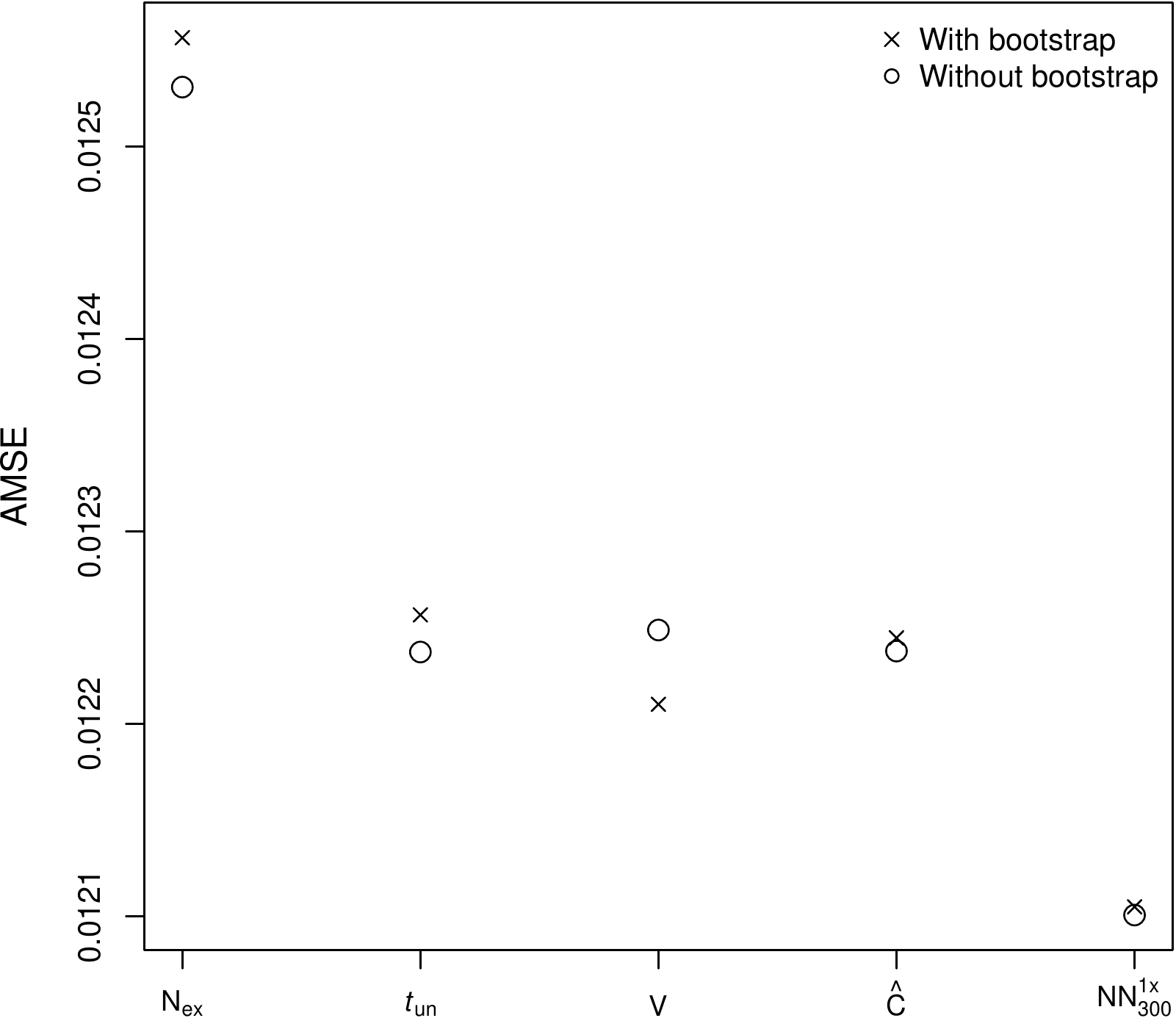}
  \hspace{1.2cm}
  \includegraphics[width=0.32\textwidth]{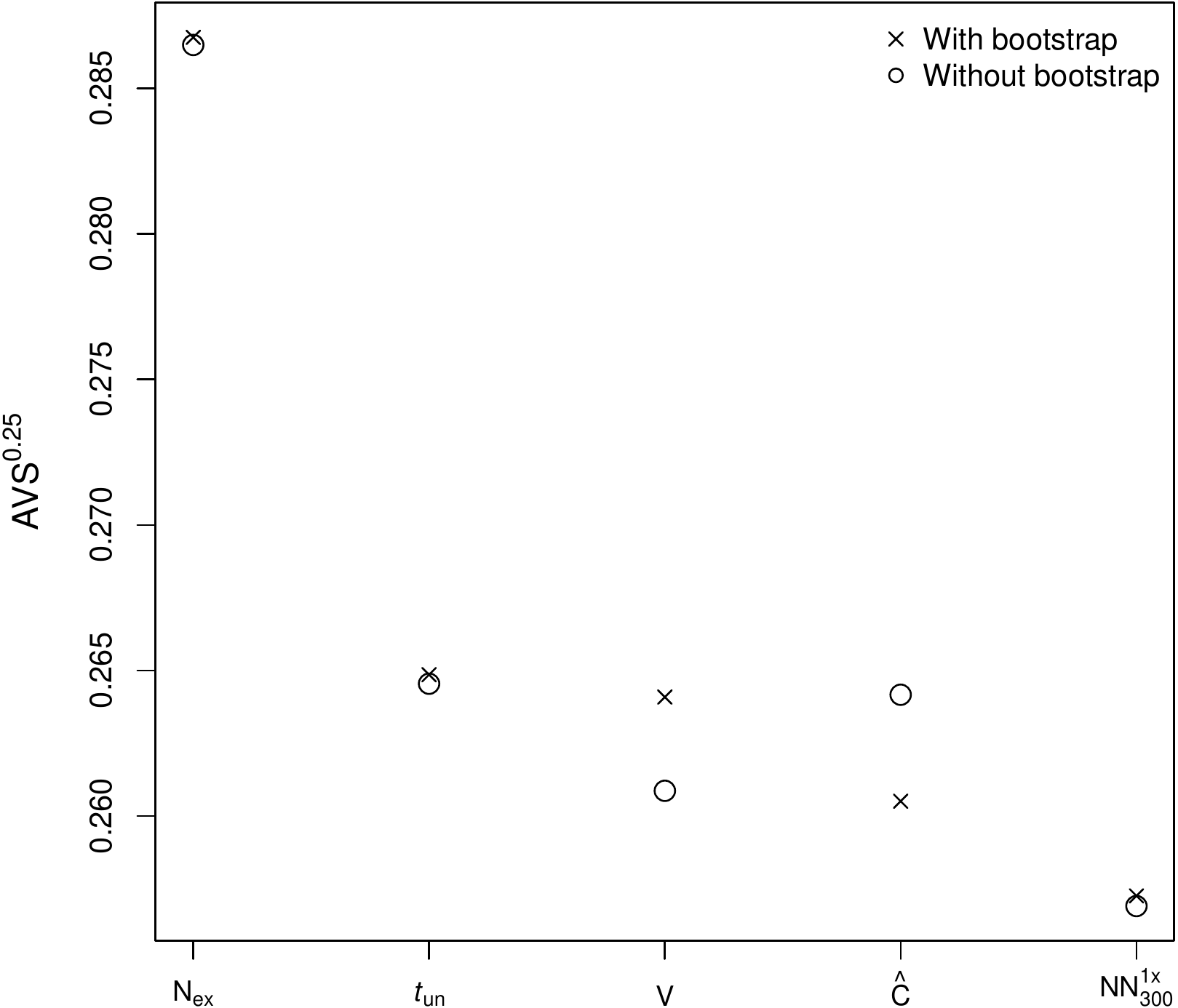}
  \caption{Scatter plots of $\AMSE$ (left column) and $\AVS^{0.25}$ (right
    column) for US (first row) and Canadian (second row) ZCB yield curve data,
    and USD (third row) and GBP (fourth row) exchange-rate data.  Each of the
    eight plots shows a comparison of the empirical predictive distributions
    produced with (using $n_{\text{bt}}=100$) and without the bootstrap. We see
    that for every considered dependence model, both methods produce fairly
    similar quality of empirical predictive distributions according to the two
    metrics. Moreover, the GMMN models produce the best empirical predictive
    distributions across all methods and datasets.}\label{fig:boot:exchange:interest}
\end{figure}

Using $n_{\text{bt}}=100$, we conducted the bootstrap experiments on the same
data sets from Section~\ref{sec:applications} for a selected subset of
dependence models with varying performance levels --- specifically, the
single-hidden-layer NN with 300 neurons per layer
($\text{NN}^{1\text{x}}_{300}$), the normal copula with exchangeable correlation
matrix ($\text{N}_{\text{ex}}$), the $t$ copula with unstructured correlation
matrix ($t_{\text{un}}$), the vine copula (V), and the empirical copula
($\hat{C}$).
We compared the quality of the empirical predictive distributions
produced with and without the bootstrap.
Figure~\ref{fig:boot:exchange:interest} displays scatter plots of the $\AMSE$
(left column) and $\AVS^{0.25}$ metrics for US ZCB yield curves (first row),
Canadian ZCB yield curves (second row), USD exchange-rate data (third row) and
GBP exchange-rate data (fourth row).
For any given dependence model, the empirical predictive distributions produced
with and without the bootstrap showed similar performances. With the
bootstrap, the GMMN continued to perform better than the other dependence
models.

Our $\AMSE$ and $\AVS^{0.25}$ metrics are overall assessments of
one-day-ahead empirical predictive distributions, averaged across all days $t = \tau+1, \dots, T$ in the test period.
We can also define corresponding daily metrics $\text{MSE}(t)$ and
$\text{VS}(t)$ so that our equations \eqref{eq:average:MSE} and
\eqref{eq:average:VS} amount to
$\text{AMSE}=(T-\tau)^{-1}\sum_{t=\tau+1}^T \text{MSE}(t)$ and
$\text{AVS}^r=(T-\tau)^{-1}\sum_{t=\tau+1}^T \text{VS}^r(t)$, respectively, and
assess the daily performance ratios with and without the bootstrap, i.e.,
$\text{MSE}_{\text{bt}}(t)/\text{MSE}_{\text{nbt}}(t)$ and
$\text{VS}^{0.25}_{\text{bt}}(t)/\text{VS}^{0.25}_{\text{nbt}}(t)$, where the
subscript ``nbt'' means ``no bootstrap''.  Over the entire test period
$t=\tau+1,\dots,T$, both these ratios turn out to fluctuate little around one across
all models and datasets considered.  Thus, our empirical investigation here
suggests that accounting for the uncertainty in the training of cross-sectional
dependence models appears to have little impact on the quality of the final
probabilistic forecasts produced.

\Urlmuskip=0mu plus 1mu\relax%
\printbibliography[heading=bibintoc]
\end{document}

%
%
%
%
